\documentclass[english,aps,prl,twocolumn,amsmath,amssymb,showpacs,superscriptaddress,notitlepage,longbibliography]{revtex4-1}
\usepackage[T1]{fontenc}
\usepackage[latin9]{inputenc}
\usepackage{color}
\usepackage{bm}
\usepackage{amsmath}
\usepackage{amssymb}
\usepackage{graphicx}
\usepackage{wasysym}

\makeatletter

\usepackage{amssymb}
\usepackage{amsmath}
\usepackage{graphicx}
\usepackage[colorlinks=true,linkcolor=blue,anchorcolor=red,citecolor=blue,urlcolor=blue]{hyperref}
\usepackage[caption=false]{subfig}
\usepackage{lipsum}

\makeatother

\usepackage{babel}
\begin{document}
\renewcommand{\figurename}{Fig.}
\title{Higher-order Weyl superconductors with anisotropic Weyl-point connectivity}
\author{W.\ B. Rui}
\email{wenbin.rui@gmail.com}

\address{Department of Physics and HKU-UCAS Joint Institute for Theoretical
and Computational Physics at Hong Kong, The University of Hong Kong,
Pokfulam Road, Hong Kong, China}
\author{Song-Bo Zhang}
\email{songbo.zhang@physik.uni-wuerzburg.de}

\address{Institute for Theoretical Physics and Astrophysics, University of
W\"urzburg, D-97074 W\"urzburg, Germany}
\author{Moritz M. Hirschmann}
\address{Max-Planck-Institute for Solid State Research, Heisenbergstrasse 1,
D-70569 Stuttgart, Germany}
\author{Andreas P. Schnyder}
\address{Max-Planck-Institute for Solid State Research, Heisenbergstrasse 1,
D-70569 Stuttgart, Germany}
\author{Bj\"{o}rn Trauzettel}
\address{Institute for Theoretical Physics and Astrophysics, University of
W\"urzburg, D-97074 W\"urzburg, Germany}
\author{Z.\ D. Wang}
\email{zwang@hku.hk}

\address{Department of Physics and HKU-UCAS Joint Institute for Theoretical
and Computational Physics at Hong Kong, The University of Hong Kong,
Pokfulam Road, Hong Kong, China}
\date{\today}
\begin{abstract}
Weyl superconductors feature Weyl points at zero energy in the three-dimensional
(3D) Brillouin zone and arc states that connect the projections of
these Weyl points on the surface. We report that higher-order Weyl
superconductors can be realized in odd-parity topological superconductors
with time-reversal symmetry being broken by periodic driving. Different
from conventional Weyl points, the higher-order Weyl points in the
bulk separate 2D first- and second-order topological phases, while
on the surface, their projections are connected
not only by conventional surface Majorana arcs, but also by hinge
Majorana arcs. We show that the
Weyl-point connectivity via Majorana arcs is largely enriched by the
underlying higher-order topology and becomes anisotropic with respect
to surface orientations. We identify the anisotropic Weyl-point connectivity
as a characteristic feature of higher-order Weyl materials. As each
2D subsystem can be singled out by fixing the periodic driving, we
propose how the Majorana zero modes in the 2D higher-order topological
phases can be detected and manipulated in experiments.
\end{abstract}
\maketitle
\textit{\textcolor{blue}{Introduction.}}\textit{\textemdash }The particular
excitations of topological semimetals and nodal superconductors emerge
around gapless degeneracies and constitute one of the main research
activities in the field of topological materials \citep{Chiu_RMP_2016,Volovik-book,zhao_PRL_2013,sato_majorana_2016,Sato_Topological_2017,Armitage2018}.
Typical examples are Weyl and Dirac semimetals or nodal superconductors
whose low-energy physics around the gapless points can be described
by Weyl or Dirac Hamiltonians \citep{Wan_PRB_2011,xu_discovery_2015,liu_stable_2014,Meng_PRB_2012,Yang_PRL_2014,Rui_PT_2019,Morali_magnetic_Weyl_2019,Liu_magnetic_Weyl_2019}.
Besides exotic quasiparticles in the bulk, the bulk topology of the
systems also gives rise to fascinating topological boundary states.
Conventionally, in an $n$-dimensional topological phase, the topological
boundary states are constrained to $(n-1)$ dimensions.

Recently, inspired by higher-order topology \citep{Benalcazar61,song_PRL_2017,Piet_PRL_2017,Peterson_fractional_2020,Schindl18science,Schindler2018higher,serra2018observation,Ezawa18PRL,Trifunovic19PRX,ChenR20PRL,Eslam_PRB_2018},
new topological phases, termed higher-order topological gapless phases,
have attracted increasing interest \citep{Lin_PRB_2018,zhang_2019_higherorder,wang_2020_higherorder,ghorashi_second-order_2019,ghorashi_2020_higherorder,wang2020boundary}.
In addition to gapless degeneracies in the $n$-dimensional bulk and
conventional $(n-1)$-dimensional boundary states, these topological
systems feature also $(n-d)$-dimensional hinge or corner states with
$d\geqslant2$. As an important member of gapless
phases, Weyl superconductors must break time-reversal or inversion
symmetry \citep{Meng_PRB_2012,Cho12PRB,Yang_PRL_2014,Bednik15PRB}.
Time-reversal-symmetry breaking is particularly important, as most
reported first-order (conventional) Weyl superconductors are realized
in this way \citep{Biswas_PRB_2013,Fischer_PRB_2014,goswami_2013_topological,hayes_2020_weyl,wang_evidence_2020,yamashita_colossal_2015,Yanase_2017_PRB,Yanase_PRB_2016,Yuan_PRB_2017}.
However, so far, there has been no study on higher-order Weyl superconductors
(HOWSCs) with broken time-reversal symmetry. An interesting fundamental
question in HOWSCs concerns their boundary states, namely, how the
connectivity of surface projected Weyl points by Majorana arcs, a
characteristic feature of Weyl superconductors, is reshaped by the
higher-order topology.

In this Letter, we show that HOWSCs with broken time-reversal symmetry
can be realized by periodically driving a 2D second-order odd-parity
topological superconductor. The periodic driving breaks time-reversal
symmetry and offers an unprecedented way to extend the 2D superconductor
to a third dimension with periodic boundary conditions. Weyl points
can be generated in this dynamic process, which split the system into
different regions of first- (FOTP) or second-order topological phases
(SOTP), leading to a HOWSC. In sharp contrast to
the surface Majorana arcs protected by a bulk Chern number in the
FOTP regions, the hinge Majorana arcs in the SOTP regions, which are
protected by inversion symmetry, depend strongly on the surface orientation
due to the higher-order topology. This results in an intriguing and
diverse recombination of surface and hinge Majorana arcs upon orientation
change, leading to an anisotropic Weyl-point connectivity. By developing
an effective boundary theory capable of describing both surface and
hinge Majorana arcs, we thoroughly analyze this intricate Weyl-point
connectivity of Majorana arcs in every surface orientation. Furthermore,
both the FOTP and SOTP can be individually investigated as the periodic
driving offers an advantage to single out each 2D slice of the system
by fixing the driving parameters. We propose to control and detect
the Majorana zero modes in the SOTP regions via circularly polarized
light (CPL) in experiments.

\begin{figure}
\includegraphics[width=1\columnwidth]{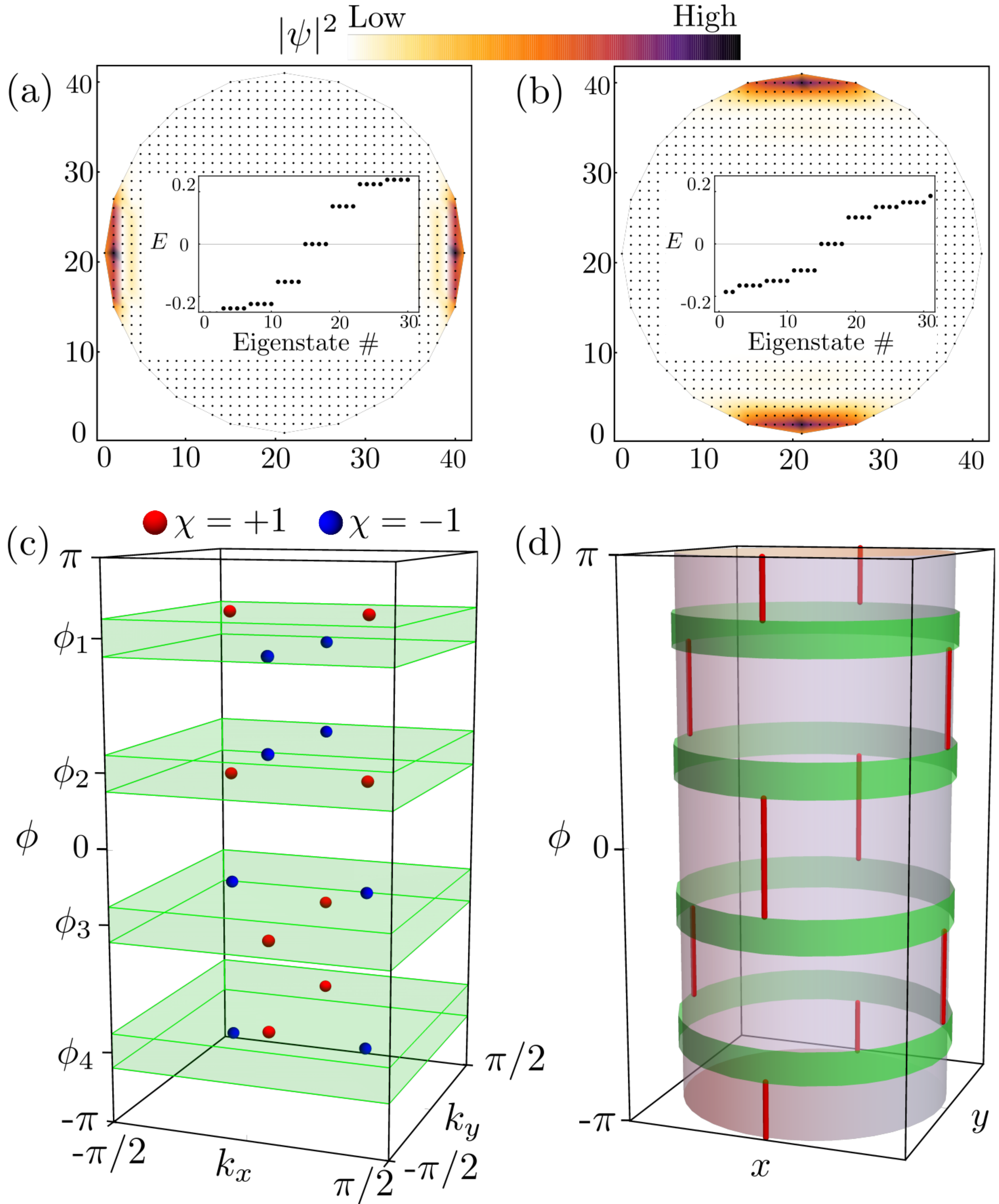}

\caption{(a) Two Majorana Kramers pairs (indicated by dark densities) are localized
at a disk boundary in the horizontal direction; (b) Circularly
polarized light with intensity $\mathcal{\mathcal{I}}=1.5(\omega/2m)$
and $\phi=0$ switches the Majoranas to the vertical direction. The
insets in (a) and (b) show the corresponding energy levels. (c) The
16 Weyl points of the model (\ref{eq:ham_modified}) in the synthetic
3D Brillouin zone. The red and blue points denote Weyl points with
positive and negative chirality $\chi=\pm1$, respectively. The Weyl
points are grouped into four sets, denoted by their center positions
$\phi_{j}$ with $j\in\{1,2,3,4\}$. Each set consists of four Weyl
points. Their positions divide the Brillouin zone into different topological
sectors. We identify FOTPs (green area) within each of the four sets
and SOTPs between neighboring sets. (d) Schematic of boundary states,
obtained by stacking the 2D disks in the third dimension parameterized
by $\phi$. In the FOTP regions, the boundary states circle around
the whole disk boundary (green belts), whereas in the SOTP regions,
they form hinge states on 1D lines (red lines). The
parameters are $M_{0}=3,m=1.1,v=1,\mu=0,\omega=0.1$, and $\Delta_{0}=0.6$.
\label{fig:1}}
\end{figure}

\textit{\textcolor{blue}{Realization of HOWSCs.}}\textit{\textemdash }Our
starting point is a 2D second-order odd-parity topological superconductor
that respects time-reversal symmetry. Different from previous proposals
for higher-order topogical superconductors \citep{CHHsu18PRL,LiuT18PRB,YanZB18PRL,WangQY18PRL,ZhuXY18PRB,zhu_PRL_2019,Geier18PRB,yan_PRL_2019,Volpez19PRL,RXZhang19PRL,ZhangSB20PRR,Plekhanov19PRR,WUYJ20PRL,PanXH19PRL,Yang_PRR_2020,ZhangSB20arxiv2,ZhangSB20arxiv1,Tiwari20PRL,Laubscher20PRR,HsuYT20PRL,XXWu19arXiv190510648W,Peng20PRR,Bomantara20PRB},
we consider an inter-orbital $s$-wave pairing potential with a constant
magnitude $\Delta_{0}$, which may be induced via the proximity effect.
The minimal Hamiltonian in momentum space can be written as $\mathcal{H}=\mathcal{H}_{0}+h_{\Delta}$
with
\begin{align}
\mathcal{H}_{0} & =m({\bf k})\tau_{z}\sigma_{z}+v\sin k_{x}s_{z}\sigma_{x}+v\sin k_{y}\tau_{z}\sigma_{y}-\mu\tau_{z},\nonumber \\
h_{\Delta} & =\Delta_{0}\tau_{y}s_{y}\sigma_{x},\label{eq:ham}
\end{align}
where $m(\mathbf{k})=M_{0}-2m(\cos k_{x}+\cos k_{y})$ and the Pauli
matrices ${\bf s}$, $\bm{{\bf \sigma}}$ and $\bm{\tau}$ act on
spin, orbital, and Nambu spaces, respectively. $\mu$ is the chemical
potential, $M_{0}$, $m$ and $v$ are material dependent parameters.
The Hamiltonian is invariant under time-reversal ($\mathcal{T}$)
and particle-hole ($\mathcal{C}$) symmetry. The pairing interaction
is of odd parity as indicated by $\mathcal{P}h_{\Delta}\mathcal{P}^{-1}=-h_{\Delta}$
with the inversion operator $\mathcal{P=\sigma}_{z}$. Correspondingly,
the BdG Hamiltonian is symmetric under inversion $\tilde{\mathcal{P}}\mathcal{H}(\mathbf{k})\tilde{\mathcal{P}}^{-1}=\mathcal{H}(\mathbf{-k})$
with $\tilde{\mathcal{P}}=\tau_{z}\mathcal{P}$. Furthermore, spin
rotation about the $z$ axis $J_{z}=\tau_{z}s_{z}$ is preserved.
Due to the second-order topology, our model features two 0D Majorana
Kramers pairs at a disk boundary in the horizontal direction, which
are protected by time-reversal and inversion symmetries {[}Fig.\ \ref{fig:1}(a){]}.

Next, we show that HOWSCs can be generated on the basis of the model
(\ref{eq:ham}) through periodic driving. For concreteness, we consider
periodic driving in the form of CPL which is shed on the system in
the $z$ direction and described by the vector potential $\mathbf{A}(t)=A_{0}(\cos(\omega t),\sin(\omega t+\phi),0)$.
$\phi$ characterizes the phase shift, $A_{0}$ the strength and $\omega$
the frequency. The CPL couples to the electrons(holes) via the Peierls
substitutions ${\bf k}\rightarrow{\bf k}\pm e{\bf A}(t)$. To proceed
analytically and elucidate our main results, we employ Floquet theory
and derive a static effective Hamiltonian \citep{SuppInf}. The effective
Hamiltonian is obtained on the basis of Eq.\ (\ref{eq:ham}) and
contains a non-trivial correction that preserves spin-rotation symmetry
about the $z$ axis. We can find it as \citep{SuppInf}
\begin{equation}
h({\bf k})=h_{0}({\bf k})+\gamma(\mathbf{k})\cos\phi,\label{eq:ham_modified}
\end{equation}
where $\gamma(\mathbf{k})=(2m\mathcal{I}/\omega)(v\sin k_{x}\sigma_{x}+v\sin k_{y}\sigma_{y}-v^{2}\sigma_{z}/2m)$
and $h_{\text{0}}({\bf k})=\tau_{z}[(m(\mathbf{k})+m\mathcal{I})\sigma_{z}+v\sin k_{x}\sigma_{x}+v\sin k_{y}\sigma_{y}-\mu]-\Delta_{0}\tau_{x}\sigma_{x}$
with $\mathcal{I}=e^{2}A_{0}^{2}$ corresponding to the intensity
of the light. The periodic driving breaks time-reversal symmetry.
Increasing $\mathcal{I}$ above a critical value, we observe that
the Majorana zero modes at the disk boundary jump from the horizontal
to vertical positions {[}Fig.\ \ref{fig:1}(b){]}.

The model in Eq.\ (\ref{eq:ham_modified}) is periodic in the parameter
$\phi$. We may regard it as an extra (third) dimension. Since
at each $\phi$ time-reversal symmetry is broken, the 2D systems for
fixed $\phi$ belong to class A and are characterized by a Chern number
\citep{Schnyder_PRB_2008}. Strikingly, stacking these 2D systems
along the $\phi$ direction gives a 3D Weyl superconductor with 16
Weyl points in the synthetic 3D Brillouin zone, as displayed in Fig.\ \ref{fig:1}(c).
These Weyl points can be grouped into four distinct sets.

In Fig.\ \ref{fig:1}(d), we stack the 2D disks with different $\phi$,
forming a 3D cylinder. The cylinder is finite in $x$ and $y$ directions
but periodic in $\phi$ direction. As can be seen by the dimensions
of the boundary states, the system splits into two kinds of topological
phases: (i) FOTPs within each of the Weyl-point sets, with 2D surface
states at the boundary (green belts); (ii) SOTPs between different
Weyl-point sets, with 1D hinge states (red lines). As the Weyl points
mediate between the FOTP and SOTPs, we coin the system a HOWSC.

\begin{figure*}[t]
\includegraphics[width=1\textwidth]{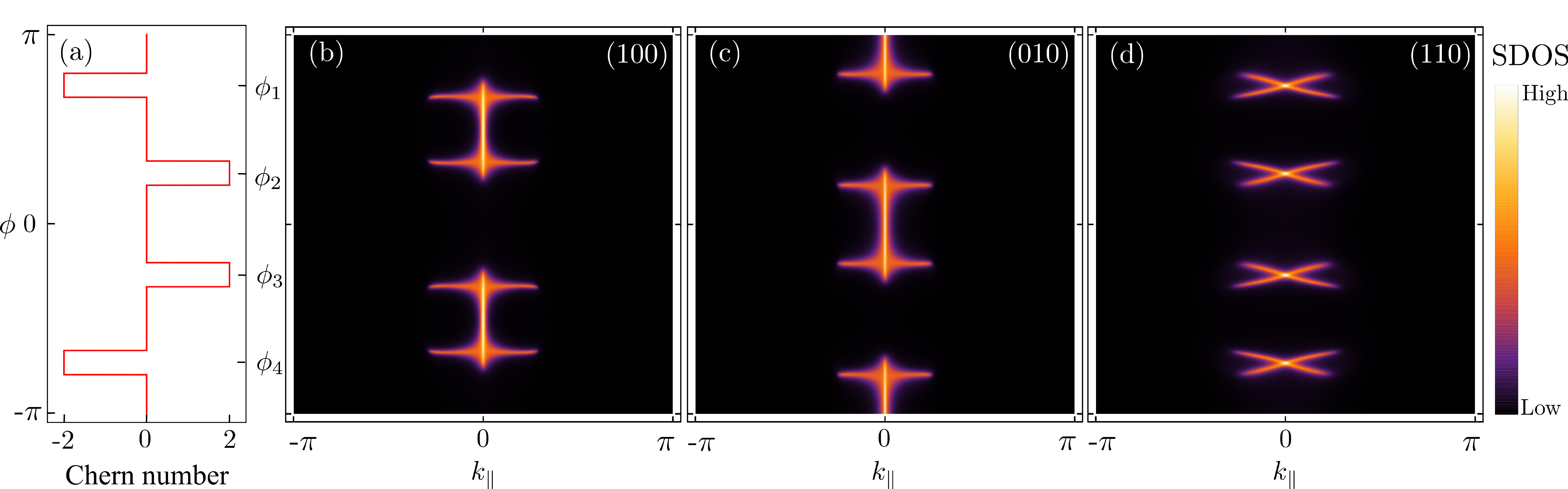}

\caption{Anisotropic Weyl-point connectivity due to the higher-order topology.
(a) Chern number calculated within the $k_{x}k_{y}$-planes as a function
of $\phi$. (b)-(d) Surface density of states on the (100), (010)
and (110) surfaces, respectively. Surface Majorana arcs connect Weyl
points inside each Weyl-point set and form horizontal bars. In contrast,
the two hinge Majorana arcs connect Weyl points from two neighboring
sets and form vertical bars, which depend strongly on the respective
surface orientation. (b) On the (100) surface, the hinge Majorana
arcs (vertical bars) connect the Weyl points between $\text{\ensuremath{\phi_{1}}(\ensuremath{\phi_{3}})}$
and $\text{\ensuremath{\phi_{2}(\phi_{4})}}$, and form two rotated
``H'' shapes; (c) On the (010) surface, the hinge Majorana arcs
connect the Weyl points between $\text{\ensuremath{\phi_{4}}(\ensuremath{\phi_{2}})}$and
$\text{\ensuremath{\phi_{1}(\phi_{3})}}$; (d) On the (110) surface,
the Weyl points are only connected by surface Majorana arcs in cross
shapes.\label{fig:Fig2}}
\end{figure*}

\textit{\textcolor{blue}{Anisotropic Weyl-point connectivity.}}\textit{\textemdash }In
conventional Weyl superconductors, the surface Majorana arcs that
connect the projections of the Weyl points in the surface Brillouin
zone are protected by a non-zero Chern number. Figure\ \ref{fig:Fig2}(a)
shows the Chern number calculated in $k_{x}k_{y}$-planes for different
$\phi$ in the HOWSC. It takes the nontrivial value of 2 or $-2$
inside each set of Weyl points (corresponding to the FOTP regions).
In contrast, it vanishes between neighboring sets (corresponding to
the SOTP regions). This can be understood from the fact that there
is an equal number of Weyl points of opposite chirality in each Weyl-point
set, rendering the Chern number non-zero only inside each set. Thus,
the surface Majorana arcs in the FOTPs are protected by a Chern number,
while the hinge Majorana arcs \citep{footnote} in the SOTPs are not.
In the following, we analyze the Weyl-point connectivity via these
two different kinds of Majorana arcs.

To visualize the Weyl-point connectivity, we calculate the surface
density of states (SDOS) for different surface orientations. Remarkably,
the form of the Weyl-point connectivity is strongly anisotropic with
respect to the surface orientation. Three typical
cases of (100), (010), and (110) surfaces are displayed in Figs.\ \ref{fig:Fig2}(b)-(d),
respectively. In Fig.\ \ref{fig:Fig2}(b),
for the (100) surface, the connectivity exhibits two separated ``H''
shapes rotated by 90 degrees. In this case, while the surface Majorana
arcs form the horizontal bars of the ``H'' shapes within each Weyl
set, the hinge Majorana arcs form the vertical bars and connect the
Weyl-point sets $\phi_{1}$($\phi_{3}$) and $\phi_{2}$($\phi_{4}$). Notably, there are double hinge Majorana arcs connecting
two pairs of Weyl points with opposite chirality. Next, we turn to the (010) surface {[}Fig.\ \ref{fig:Fig2}(c){]}.
Although the connectivity still forms two rotated ``H'' shapes,
the vertical bars composed of hinge Majorana arcs now connect different
pairs of Weyl-point sets, namely $\phi_{4}$($\phi_{2}$) and $\phi_{1}$($\phi_{3}$).
Finally, in Fig.\ \ref{fig:Fig2}(d) for the (110) surface, the Weyl
points can only be connected by surface Majorana arcs in cross shapes. While surface Majorana arcs can always be observed,
the hinge Majorana arcs depend sensitively on surface orientation.
Thus, the Weyl-point connectivity is anisotropic.

\textit{\textcolor{blue}{Effective boundary theory.}}\textit{\textemdash }For
a better understanding of the orientation-dependent
connectivity of the Majorana arcs, it is instructive to develop a
boundary theory applicable to any surface orientation. To do so, we
first derive two boundary states $(\Psi_{e\uparrow},\Psi_{h\downarrow})$
for each $\phi$ in the absence of pairing interactions \citep{SuppInf}.
Using these boundary states as a basis, the resulting effective boundary
Hamiltonian can be obtained as
\begin{equation}
h_{\text{eff}}(\theta)=\left(\begin{array}{cc}
|v^{+}|k_{\parallel}-\mu & \tilde{\Delta}(\theta)\\
\tilde{\Delta}(\theta)^{*} & -|v^{-}|k_{\parallel}+\mu
\end{array}\right),\label{eq:projected_hamil}
\end{equation}
where $v^{\pm}=v(1\pm2m\mathcal{I}\cos\phi/\omega)$, $\theta$ is
the angle between the boundary and $x$ direction, and $k_{\parallel}$
is the momentum along the boundary (see Fig.\ 1 in the Supplemental
Material \citep{SuppInf}). The projected pairing potential $\tilde{\Delta}(\theta)$
is obtained as
\begin{align}
\tilde{\Delta}(\theta) & =\frac{i}{2}\mathcal{F}\Delta_{0}\textrm{sgn}(v^{-})\left(\textrm{sgn}(v^{+}v^{-})e^{i\theta}-e^{-i\theta}\right),\label{eq:mass_term}
\end{align}
with $\text{sgn}(\cdot)$ being the sign function. The prefactor $\mathcal{F}$
stems from the overlap of the boundary state wavefunctions \citep{SuppInf}.
It is unity for $\cos\phi=0$ but smaller than one in general. The
eigenenergies are given by $E_{\text{eff}}=(|v^{+}|-|v^{-}|)k_{\parallel}\ensuremath{/2+\{[(|v^{+}|+|v^{-}|)k_{\parallel}/2-\mu]^{2}+|\tilde{\Delta}(\theta)|^{2}\}^{1/2}.}$
The chemical potential $\mu$ can be absorbed in $k_{\parallel}$
in the square root and the band gap is given by $2|\tilde{\Delta}(\theta)|$.
For simplicity, we set $\mu$ to zero in the following discussion.
Notably, Eq.\ (\ref{eq:projected_hamil}) takes the form of a 1D
Dirac Hamiltonian with a Dirac mass $\tilde{\Delta}(\theta)$. The
mass gaps out the boundary spectrum everywhere, except for isolated
values of $\theta$ where $\tilde{\Delta}(\theta)=0$. This is the
reason why the appearance of hinge Majorana arcs depends sensitively
on the surface orientation in the SOTPs.

The periodic driving preserves inversion symmetry of the system. Thus,
Eq.\ (\ref{eq:projected_hamil}) obeys $\bar{\mathcal{P}}h_{\text{eff}}(\theta)\bar{\mathcal{P}}^{-1}=h_{\text{eff}}(\theta+\pi)$
with $\bar{\mathcal{P}}=\text{\ensuremath{\sigma_{z}}}$ the projected
inversion operator, enforcing a constraint on $\tilde{\Delta}(\theta)$:
$\tilde{\Delta}(\theta+\pi)=-\tilde{\Delta}(\theta).$ Obviously,
$\tilde{\Delta}(\theta)$ changes sign when advancing from $\theta$
to $\theta+\pi$, leading to a gapless point along $\theta$. The
gapless point corresponds to the positions of a hinge Majorana arc.
In this regard, the SOTP is protected by inversion symmetry. This
result is not restricted to a specific geometry as long as inversion
symmetry is preserved. From Eq.\ (\ref{eq:mass_term}), we can determine
the positions of the gapless points explicitly,
\begin{equation}
\theta=\pi[1-\textrm{sgn}(v^{+}v^{-})]/4+n\pi,\ \ n\in\{0,1\}.\label{eq:gapless_point}
\end{equation}
When $2m\mathcal{I}/\omega>1$, $v^{+}v^{-}$ changes sign at $\phi=\phi_{j}$
with $j\in\{1,2,3,4\}$, $\phi_{1}=-\phi_{4}=\pi-\arccos(\omega/2m\mathcal{I})$
and $\phi_{2}=-\phi_{3}=\arccos(\omega/2m\mathcal{I})$. As a result,
the positions in Eq.\ (\ref{eq:gapless_point}) switch from $\{0,\pi\}$
to $\{\pi/2,3\pi/2\}$.

Facilitated by the boundary theory, we are now able to explain the
anisotropic connectivity of the Majorana arcs obtained numerically
in Fig.\ \ref{fig:Fig2}. First, for the $(100)$ surface, $k_{\parallel}=k_{y}$
and $\theta=0$. In this case, the vanishing of the mass $\tilde{\Delta}(\theta)$
in Eq.\ (\ref{eq:mass_term}) is determined by $\textrm{sgn}(v^{+}v^{-})=+1$,
which gives $\phi\in(\phi_{1},\phi_{2}]\cup(\phi_{3},\phi_{4}]$.
Thus, the hinge Majorana arcs connect the Weyl-point sets $\text{\ensuremath{\phi_{1}}}$
with $\phi_{2}$ and $\phi_{3}$ with $\phi_{4}$ at $k_{y}=0$, as
shown by the vertical bars in Fig.\ \ref{fig:Fig2}(b). Second, for
the $(010)$ surface with $k_{\parallel}=k_{x}$ ($\theta=\pi/2$),
$\tilde{\Delta}(\theta)$ vanishes at $\textrm{sgn}(v^{+}v^{-})=-1$,
leading to $\phi\in(\phi_{4},\phi_{1}]\cup(\phi_{2},\phi_{3}]$. In
this orientation, the hinge Majorana arcs instead connect $\text{\ensuremath{\phi_{4}}}$
with $\phi_{1}$ and $\phi_{2}$ with $\phi_{3}$ at $k_{x}=0$, as
shown by the vertical bars in Fig.\ \ref{fig:Fig2}(c). In contrast,
for the $(110)$ surface associated with $\theta=\pi/4$, $\tilde{\Delta}(\theta)$
is always non-zero for all $\phi$ {[}Fig.\ \ref{fig:Fig2}(d){]}.
As a result, there are no hinge arcs connecting the Weyl points.

\begin{figure}
\includegraphics[width=1\columnwidth]{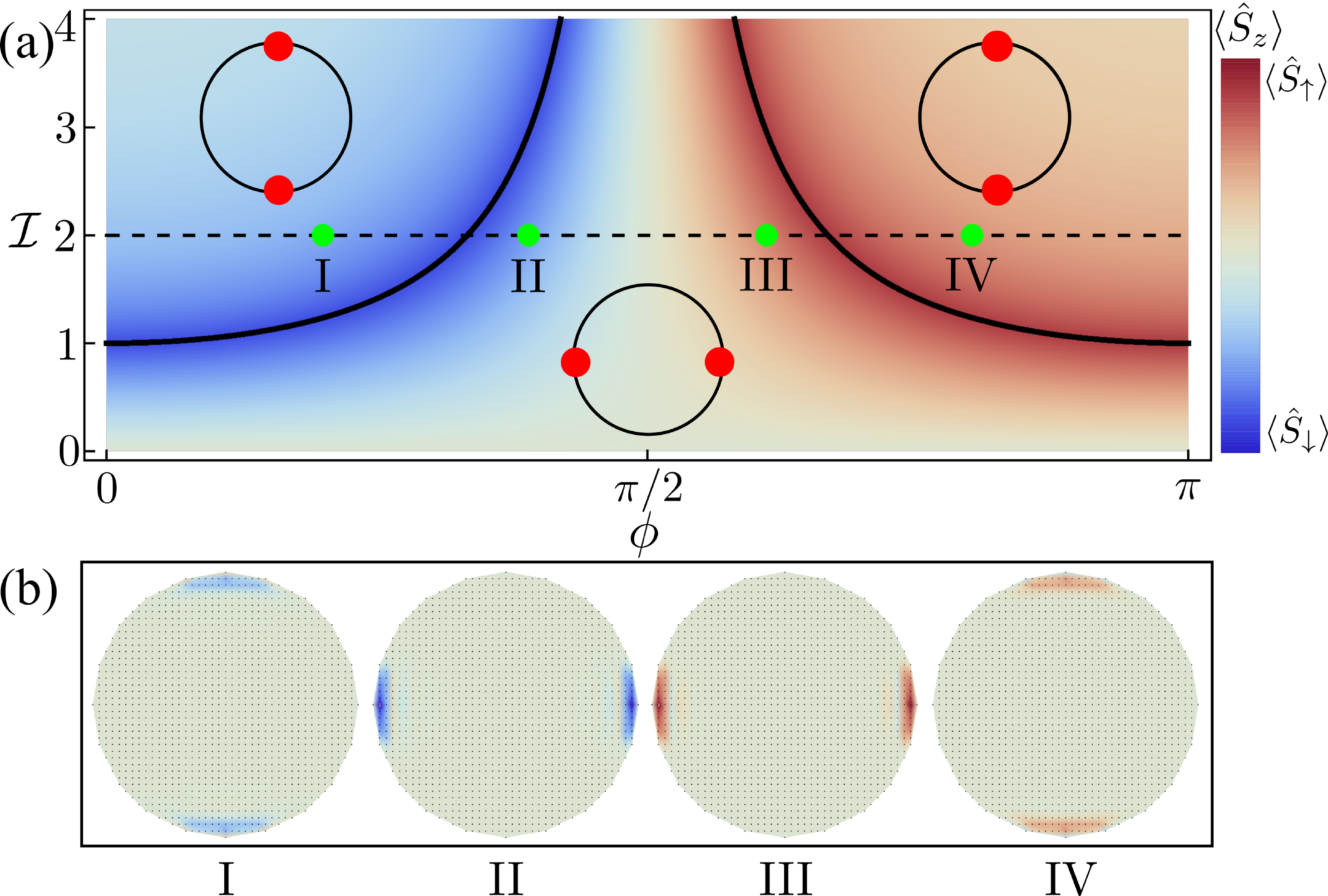}

\caption{Manipulation of Majorana zero modes in a 2D subsystem. (a) The phase
diagram for the second-order topological superconductor, colored by
spin polarization calculated with Eq.\ (\ref{eq:spin_polarization}).
The solid lines close to the positions of Weyl points separate different
SOTPs. The locations of the Majorana zero modes are sketched by the
insets. The four disks in (b) show the positions and spin polarizations
of the Majorana zero modes corresponding to the points at I-IV on
the dashed line with $\mathcal{I}=2$ in (a), obtained by tight-binding
calculations. $\mathcal{I}$ is given in units of
$\omega/(2m)$, and all other parameter values are the same as in
Fig.\ \ref{fig:1}. \label{fig:Fig3}}
\end{figure}

\textit{\textcolor{blue}{Manipulating Majorana zero modes.}}\textit{\textemdash }The
realization of detectable and tunable Majorana zero modes is one of
the main research goals in Majorana physics \citep{Kitaev_2001,Alicea_2012,Mourik1003,Lutchyn_PRL_2010,Aasen_2016_PRX,sun_PRL_2016}.
As shown in Fig.\ \ref{fig:1}, by varying $\phi$ in the SOTPs,
the positions of the Majorana zero modes at the boundary of the 2D
disk are tunable. Besides this, we find that the Majorana zero modes
has finite spin polarization which can also be controlled by tuning
CPL.

To elucidate this manipulation, we calculate explicitly
the wavefunctions of the Majorana zero modes and hence their spin
polarizations. Starting with the boundary Hamiltonian Eq.\ (\ref{eq:projected_hamil}),
we first obtain the wavefunctions of the zero-energy modes. From these
zero-energy modes, two Majorana zero modes can then be derived. Their
wavefunctions in the Nambu and spin basis $(\Psi_{e\uparrow},\Psi_{e\downarrow},$
$\Psi_{h\uparrow},\Psi_{h\downarrow})$ can be written as $\Phi_{1}\propto(e^{i\theta}\textrm{sgn}(v^{+})\sqrt{|v^{-}|},-i\sqrt{|v^{+}|},e^{-i\theta}\textrm{sgn}(v^{+})\sqrt{|v^{-}|},i\sqrt{|v^{+}|})^{T}$
and $\Phi_{2}\propto(-ie^{i\theta}\textrm{sgn}(v^{+})\sqrt{|v^{-}|},\sqrt{|v^{+}|},ie^{-i\theta}\textrm{sgn}(v^{+})\sqrt{|v^{-}|},$
$\sqrt{|v^{+}|})^{T}$ \citep{SuppInf}. The angular positions $\theta$
are either $\{0,\pi\}$ or $\{\pi/2,3\pi/2\}$, depending on $\phi$,
as we have shown before. The spin polarization of the Majorana zero
modes can be calculated as $\langle\hat{{\bf S}}\rangle_{j}=\langle\Phi_{j}|\hat{{\bf s}}|\Phi_{j}\rangle$,
where $j\in\{1,2\}$ and $\hat{{\bf s}}=\hbar(\tau_{0}+\tau_{z}){\bf s}/2$
\cite{Sticlet12PRL}. We find that the two Majorana zero modes have always opposite spins
in $x$- and $y$-directions. Thus, they together yield vanishing
$\langle\hat{S}_{x}\rangle=\langle\hat{S}_{y}\rangle=0$ at the boundary$.$
In contrast, for the $z$-component, we find an identical spin polarization
for all the Majorana zero modes,
\begin{equation}
\langle\hat{S}_{z}\rangle=\dfrac{\hbar}{2}\frac{|v^{-}|-|v^{+}|}{|v^{-}|+|v^{+}|},\label{eq:spin_polarization}
\end{equation}
which is independent of $\theta$.

A phase diagram for the spin polarization of the
Majorana zero modes with respect to the phase shift $\phi$ and intensity
$\mathcal{I}$ is plotted in Fig.\ \ref{fig:Fig3}(a). In contrast
to the positions of the Majorana zero modes, the spin polarization
depends continuously on the phase shift $\phi$. The spin polarization splits into spin-up and spin-down regions with
the border at $\phi=\pi/2$, because the correction $\text{\ensuremath{\gamma}}(\mathbf{k})\cos\phi$
in Eq.\ (\ref{eq:ham_modified}) induced by the CPL vanishes and
time-reversal symmetry is restored at the border. $\langle\hat{S}_{z}\rangle$
is an odd function of $\phi-\pi/2$. In the vicinity of $\phi=\pi/2$,
it grows linearly with increasing $\phi$, $\langle\hat{S}_{z}\rangle\approx(2\hbar m\mathcal{I}/\omega)(\phi-\pi/2)$.
The spin polarization approaches its maximal value in regions where
$v^{+}$ or $v^{-}$ becomes zero. These regions actually separate
different SOTPs, as indicated by solid lines in Fig.\ \ref{fig:Fig3}(a). Four representative cases in the diagram marked by (I-IV) are shown
in Fig.\ \ref{fig:Fig3}(b). We can see that for the phase shift
$\phi$ at I and II, the Majorana zero modes are spin-down polarized,
while at III and IV, they are spin-up polarized. We note that the
spin polarization is measurable by spin-polarized scanning tunneling
spectroscopy \citep{sun_PRL_2016}. This indicates that the Majorana
zero modes in our system can be manipulated and detected at the same
time.

\textit{\textcolor{blue}{Conclusions and discussions.}}\textit{\textemdash }We
have proposed to realize time-several symmetry broken HOWSCs in second-order
topological superconductors with odd-parity pairing potential by periodic
driving. We have revealed an important characteristic
feature of higher-order Weyl materials, namely, the anisotropic Weyl-point
connectivity of the surface and hinge Majorana arcs. We have further
shown the possibility to detect and manipulate the Majorana zero states
by CPL.

To generate Weyl points and realize the position switching of Majorana
zero modes, the frequency and intensity of CPL need to satisfy $2m\mathcal{I}/\omega>1$.
For typical values $m\simeq50$ eV$\cdot$$\mathring{\text{A}}{}^{2}$
(e.g., for inverted Hg(Cd)Te quantum wells \citep{Konig08JPSJ}) and
$\omega\simeq0.1$ eV, an experimentally feasible intensity of $eA_{0}=\sqrt{\mathcal{I}}\apprge0.032$
$\mathring{\text{A}}{}^{-1}$ is sufficient for our propose. Finally,
we remark that the conclusion drawn in the present
work is fundamental to generic higher-order Weyl materials, including
not only superconductors and semimetals, but also artificial systems.
Explicitly, the model in Eq.\ (\ref{eq:ham_modified}) and related
physics could also be simulated by superconducting quantum circuits
consisting of multiple qubits \citep{Tan_PRL_2019_Simulation}.

\begin{acknowledgements}This work was supported by the DFG (SPP1666
and SFB1170 ``ToCoTronics''), the W\"urzburg-Dresden Cluster of
Excellence ct.qmat, EXC2147, project-id 390858490, and the Elitenetzwerk
Bayern Graduate School on ``Topological Insulators''. \end{acknowledgements}

\bibliographystyle{apsrev4-1-etal-title}
\bibliography{Reference}

\begin{thebibliography}{77}%
\makeatletter
\providecommand \@ifxundefined [1]{%
 \@ifx{#1\undefined}
}%
\providecommand \@ifnum [1]{%
 \ifnum #1\expandafter \@firstoftwo
 \else \expandafter \@secondoftwo
 \fi
}%
\providecommand \@ifx [1]{%
 \ifx #1\expandafter \@firstoftwo
 \else \expandafter \@secondoftwo
 \fi
}%
\providecommand \natexlab [1]{#1}%
\providecommand \enquote  [1]{``#1''}%
\providecommand \bibnamefont  [1]{#1}%
\providecommand \bibfnamefont [1]{#1}%
\providecommand \citenamefont [1]{#1}%
\providecommand \href@noop [0]{\@secondoftwo}%
\providecommand \href [0]{\begingroup \@sanitize@url \@href}%
\providecommand \@href[1]{\@@startlink{#1}\@@href}%
\providecommand \@@href[1]{\endgroup#1\@@endlink}%
\providecommand \@sanitize@url [0]{\catcode `\\12\catcode `\$12\catcode
  `\&12\catcode `\#12\catcode `\^12\catcode `\_12\catcode `\%12\relax}%
\providecommand \@@startlink[1]{}%
\providecommand \@@endlink[0]{}%
\providecommand \url  [0]{\begingroup\@sanitize@url \@url }%
\providecommand \@url [1]{\endgroup\@href {#1}{\urlprefix }}%
\providecommand \urlprefix  [0]{URL }%
\providecommand \Eprint [0]{\href }%
\providecommand \doibase [0]{http://dx.doi.org/}%
\providecommand \selectlanguage [0]{\@gobble}%
\providecommand \bibinfo  [0]{\@secondoftwo}%
\providecommand \bibfield  [0]{\@secondoftwo}%
\providecommand \translation [1]{[#1]}%
\providecommand \BibitemOpen [0]{}%
\providecommand \bibitemStop [0]{}%
\providecommand \bibitemNoStop [0]{.\EOS\space}%
\providecommand \EOS [0]{\spacefactor3000\relax}%
\providecommand \BibitemShut  [1]{\csname bibitem#1\endcsname}%
\let\auto@bib@innerbib\@empty
\bibitem [{\citenamefont {Chiu}\ \emph {et~al.}(2016)\citenamefont {Chiu},
  \citenamefont {Teo}, \citenamefont {Schnyder},\ and\ \citenamefont
  {Ryu}}]{Chiu_RMP_2016}%
  \BibitemOpen
  \bibfield  {author} {\bibinfo {author} {\bibfnamefont {C.-K.}\ \bibnamefont
  {Chiu}}, \bibinfo {author} {\bibfnamefont {J.~C.~Y.}\ \bibnamefont {Teo}},
  \bibinfo {author} {\bibfnamefont {A.~P.}\ \bibnamefont {Schnyder}}, \ and\
  \bibinfo {author} {\bibfnamefont {S.}~\bibnamefont {Ryu}},\ }\bibfield
  {title} {\enquote {\bibinfo {title} {Classification of topological quantum
  matter with symmetries}}, }\href {\doibase 10.1103/RevModPhys.88.035005}
  {\bibfield  {journal} {\bibinfo  {journal} {Rev. Mod. Phys.}\ }\textbf
  {\bibinfo {volume} {88}},\ \bibinfo {pages} {035005} (\bibinfo {year}
  {2016})}\BibitemShut {NoStop}%
\bibitem [{\citenamefont {Volovik}(2003)}]{Volovik-book}%
  \BibitemOpen
  \bibfield  {author} {\bibinfo {author} {\bibfnamefont {G.~E.}\ \bibnamefont
  {Volovik}},\ }\href@noop {} {\emph {\bibinfo {title} {Universe in a helium
  droplet}}}\ (\bibinfo  {publisher} {Oxford University Press, Oxford UK},\
  \bibinfo {year} {2003})\BibitemShut {NoStop}%
\bibitem [{\citenamefont {Zhao}\ and\ \citenamefont
  {Wang}(2013)}]{zhao_PRL_2013}%
  \BibitemOpen
  \bibfield  {author} {\bibinfo {author} {\bibfnamefont {Y.~X.}\ \bibnamefont
  {Zhao}}\ and\ \bibinfo {author} {\bibfnamefont {Z.~D.}\ \bibnamefont
  {Wang}},\ }\bibfield  {title} {\enquote {\bibinfo {title} {{Topological
  Classification and Stability of Fermi Surfaces}}}, }\href {\doibase
  10.1103/PhysRevLett.110.240404} {\bibfield  {journal} {\bibinfo  {journal}
  {Phys. Rev. Lett.}\ }\textbf {\bibinfo {volume} {110}},\ \bibinfo {pages}
  {240404} (\bibinfo {year} {2013})}\BibitemShut {NoStop}%
\bibitem [{\citenamefont {Sato}\ and\ \citenamefont
  {Fujimoto}(2016)}]{sato_majorana_2016}%
  \BibitemOpen
  \bibfield  {author} {\bibinfo {author} {\bibfnamefont {M.}~\bibnamefont
  {Sato}}\ and\ \bibinfo {author} {\bibfnamefont {S.}~\bibnamefont
  {Fujimoto}},\ }\bibfield  {title} {\enquote {\bibinfo {title} {{Majorana
  Fermions and Topology in Superconductors}}}, }\href {\doibase
  10.7566/JPSJ.85.072001} {\bibfield  {journal} {\bibinfo  {journal} {Journal
  of the Physical Society of Japan}\ }\textbf {\bibinfo {volume} {85}},\
  \bibinfo {pages} {072001} (\bibinfo {year} {2016})}\BibitemShut {NoStop}%
\bibitem [{\citenamefont {Sato}\ and\ \citenamefont
  {Ando}(2017)}]{Sato_Topological_2017}%
  \BibitemOpen
  \bibfield  {author} {\bibinfo {author} {\bibfnamefont {M.}~\bibnamefont
  {Sato}}\ and\ \bibinfo {author} {\bibfnamefont {Y.}~\bibnamefont {Ando}},\
  }\bibfield  {title} {\enquote {\bibinfo {title} {Topological superconductors:
  a review}}, }\href {\doibase 10.1088/1361-6633/aa6ac7} {\bibfield  {journal}
  {\bibinfo  {journal} {Reports on Progress in Physics}\ }\textbf {\bibinfo
  {volume} {80}},\ \bibinfo {pages} {076501} (\bibinfo {year}
  {2017})}\BibitemShut {NoStop}%
\bibitem [{\citenamefont {Armitage}\ \emph {et~al.}(2018)\citenamefont
  {Armitage}, \citenamefont {Mele},\ and\ \citenamefont
  {Vishwanath}}]{Armitage2018}%
  \BibitemOpen
  \bibfield  {author} {\bibinfo {author} {\bibfnamefont {N.~P.}\ \bibnamefont
  {Armitage}}, \bibinfo {author} {\bibfnamefont {E.~J.}\ \bibnamefont {Mele}},
  \ and\ \bibinfo {author} {\bibfnamefont {A.}~\bibnamefont {Vishwanath}},\
  }\bibfield  {title} {\enquote {\bibinfo {title} {{Weyl and Dirac semimetals
  in three-dimensional solids}}}, }\href {\doibase
  10.1103/RevModPhys.90.015001} {\bibfield  {journal} {\bibinfo  {journal}
  {Rev. Mod. Phys.}\ }\textbf {\bibinfo {volume} {90}},\ \bibinfo {pages}
  {015001} (\bibinfo {year} {2018})}\BibitemShut {NoStop}%
\bibitem [{\citenamefont {Wan}\ \emph {et~al.}(2011)\citenamefont {Wan},
  \citenamefont {Turner}, \citenamefont {Vishwanath},\ and\ \citenamefont
  {Savrasov}}]{Wan_PRB_2011}%
  \BibitemOpen
  \bibfield  {author} {\bibinfo {author} {\bibfnamefont {X.}~\bibnamefont
  {Wan}}, \bibinfo {author} {\bibfnamefont {A.~M.}\ \bibnamefont {Turner}},
  \bibinfo {author} {\bibfnamefont {A.}~\bibnamefont {Vishwanath}}, \ and\
  \bibinfo {author} {\bibfnamefont {S.~Y.}\ \bibnamefont {Savrasov}},\
  }\bibfield  {title} {\enquote {\bibinfo {title} {{Topological semimetal and
  Fermi-arc surface states in the electronic structure of pyrochlore
  iridates}}}, }\href {\doibase 10.1103/PhysRevB.83.205101} {\bibfield
  {journal} {\bibinfo  {journal} {Phys. Rev. B}\ }\textbf {\bibinfo {volume}
  {83}},\ \bibinfo {pages} {205101} (\bibinfo {year} {2011})}\BibitemShut
  {NoStop}%
\bibitem [{\citenamefont {Xu}\ \emph {et~al.}(2015)\citenamefont {Xu},
  \citenamefont {Belopolski}, \citenamefont {Alidoust}, \citenamefont
  {Neupane}, \citenamefont {Bian}, \citenamefont {Zhang}, \citenamefont
  {Sankar}, \citenamefont {Chang}, \citenamefont {Yuan}, \citenamefont {Lee},
  \citenamefont {Huang}, \citenamefont {Zheng}, \citenamefont {Ma},
  \citenamefont {Sanchez}, \citenamefont {Wang}, \citenamefont {Bansil},
  \citenamefont {Chou}, \citenamefont {Shibayev}, \citenamefont {Lin},
  \citenamefont {Jia},\ and\ \citenamefont {Hasan}}]{xu_discovery_2015}%
  \BibitemOpen
  \bibfield  {author} {\bibinfo {author} {\bibfnamefont {S.-Y.}\ \bibnamefont
  {Xu}}, \bibinfo {author} {\bibfnamefont {I.}~\bibnamefont {Belopolski}},
  \bibinfo {author} {\bibfnamefont {N.}~\bibnamefont {Alidoust}}, \bibinfo
  {author} {\bibfnamefont {M.}~\bibnamefont {Neupane}}, \bibinfo {author}
  {\bibfnamefont {G.}~\bibnamefont {Bian}}, \bibinfo {author} {\bibfnamefont
  {C.}~\bibnamefont {Zhang}},  \emph {et~al.},\ }\bibfield  {title} {\enquote
  {\bibinfo {title} {{Discovery of a Weyl fermion semimetal and topological
  Fermi arcs}}}, }\href {\doibase 10.1126/science.aaa9297} {\bibfield
  {journal} {\bibinfo  {journal} {Science}\ }\textbf {\bibinfo {volume}
  {349}},\ \bibinfo {pages} {613} (\bibinfo {year} {2015})}\BibitemShut
  {NoStop}%
\bibitem [{\citenamefont {Liu}\ \emph {et~al.}(2014)\citenamefont {Liu},
  \citenamefont {Jiang}, \citenamefont {Zhou}, \citenamefont {Wang},
  \citenamefont {Zhang}, \citenamefont {Weng}, \citenamefont {Prabhakaran},
  \citenamefont {Mo}, \citenamefont {Peng}, \citenamefont {Dudin},
  \citenamefont {Kim}, \citenamefont {Hoesch}, \citenamefont {Fang},
  \citenamefont {Dai}, \citenamefont {Shen}, \citenamefont {Feng},
  \citenamefont {Hussain},\ and\ \citenamefont {Chen}}]{liu_stable_2014}%
  \BibitemOpen
  \bibfield  {author} {\bibinfo {author} {\bibfnamefont {Z.~K.}\ \bibnamefont
  {Liu}}, \bibinfo {author} {\bibfnamefont {J.}~\bibnamefont {Jiang}}, \bibinfo
  {author} {\bibfnamefont {B.}~\bibnamefont {Zhou}}, \bibinfo {author}
  {\bibfnamefont {Z.~J.}\ \bibnamefont {Wang}}, \bibinfo {author}
  {\bibfnamefont {Y.}~\bibnamefont {Zhang}}, \bibinfo {author} {\bibfnamefont
  {H.~M.}\ \bibnamefont {Weng}},  \emph {et~al.},\ }\bibfield  {title}
  {\enquote {\bibinfo {title} {{A stable three-dimensional topological Dirac
  semimetal Cd$_3$As$_2$}}}, }\href {\doibase 10.1038/nmat3990} {\bibfield
  {journal} {\bibinfo  {journal} {Nature Materials}\ }\textbf {\bibinfo
  {volume} {13}},\ \bibinfo {pages} {677} (\bibinfo {year} {2014})}\BibitemShut
  {NoStop}%
\bibitem [{\citenamefont {Meng}\ and\ \citenamefont
  {Balents}(2012)}]{Meng_PRB_2012}%
  \BibitemOpen
  \bibfield  {author} {\bibinfo {author} {\bibfnamefont {T.}~\bibnamefont
  {Meng}}\ and\ \bibinfo {author} {\bibfnamefont {L.}~\bibnamefont {Balents}},\
  }\bibfield  {title} {\enquote {\bibinfo {title} {Weyl superconductors}},
  }\href {\doibase 10.1103/PhysRevB.86.054504} {\bibfield  {journal} {\bibinfo
  {journal} {Phys. Rev. B}\ }\textbf {\bibinfo {volume} {86}},\ \bibinfo
  {pages} {054504} (\bibinfo {year} {2012})}\BibitemShut {NoStop}%
\bibitem [{\citenamefont {Yang}\ \emph {et~al.}(2014)\citenamefont {Yang},
  \citenamefont {Pan},\ and\ \citenamefont {Zhang}}]{Yang_PRL_2014}%
  \BibitemOpen
  \bibfield  {author} {\bibinfo {author} {\bibfnamefont {S.~A.}\ \bibnamefont
  {Yang}}, \bibinfo {author} {\bibfnamefont {H.}~\bibnamefont {Pan}}, \ and\
  \bibinfo {author} {\bibfnamefont {F.}~\bibnamefont {Zhang}},\ }\bibfield
  {title} {\enquote {\bibinfo {title} {{Dirac and Weyl Superconductors in Three
  Dimensions}}}, }\href {\doibase 10.1103/PhysRevLett.113.046401} {\bibfield
  {journal} {\bibinfo  {journal} {Phys. Rev. Lett.}\ }\textbf {\bibinfo
  {volume} {113}},\ \bibinfo {pages} {046401} (\bibinfo {year}
  {2014})}\BibitemShut {NoStop}%
\bibitem [{\citenamefont {Rui}\ \emph {et~al.}(2019)\citenamefont {Rui},
  \citenamefont {Hirschmann},\ and\ \citenamefont {Schnyder}}]{Rui_PT_2019}%
  \BibitemOpen
  \bibfield  {author} {\bibinfo {author} {\bibfnamefont {W.~B.}\ \bibnamefont
  {Rui}}, \bibinfo {author} {\bibfnamefont {M.~M.}\ \bibnamefont {Hirschmann}},
  \ and\ \bibinfo {author} {\bibfnamefont {A.~P.}\ \bibnamefont {Schnyder}},\
  }\bibfield  {title} {\enquote {\bibinfo {title} {{$\mathcal{PT}$-symmetric
  non-Hermitian Dirac semimetals}}}, }\href {\doibase
  10.1103/PhysRevB.100.245116} {\bibfield  {journal} {\bibinfo  {journal}
  {Phys. Rev. B}\ }\textbf {\bibinfo {volume} {100}},\ \bibinfo {pages}
  {245116} (\bibinfo {year} {2019})}\BibitemShut {NoStop}%
\bibitem [{\citenamefont {Morali}\ \emph {et~al.}(2019)\citenamefont {Morali},
  \citenamefont {Batabyal}, \citenamefont {Nag}, \citenamefont {Liu},
  \citenamefont {Xu}, \citenamefont {Sun}, \citenamefont {Yan}, \citenamefont
  {Felser}, \citenamefont {Avraham},\ and\ \citenamefont
  {Beidenkopf}}]{Morali_magnetic_Weyl_2019}%
  \BibitemOpen
  \bibfield  {author} {\bibinfo {author} {\bibfnamefont {N.}~\bibnamefont
  {Morali}}, \bibinfo {author} {\bibfnamefont {R.}~\bibnamefont {Batabyal}},
  \bibinfo {author} {\bibfnamefont {P.~K.}\ \bibnamefont {Nag}}, \bibinfo
  {author} {\bibfnamefont {E.}~\bibnamefont {Liu}}, \bibinfo {author}
  {\bibfnamefont {Q.}~\bibnamefont {Xu}}, \bibinfo {author} {\bibfnamefont
  {Y.}~\bibnamefont {Sun}}, \bibinfo {author} {\bibfnamefont {B.}~\bibnamefont
  {Yan}}, \bibinfo {author} {\bibfnamefont {C.}~\bibnamefont {Felser}},
  \bibinfo {author} {\bibfnamefont {N.}~\bibnamefont {Avraham}}, \ and\
  \bibinfo {author} {\bibfnamefont {H.}~\bibnamefont {Beidenkopf}},\ }\bibfield
   {title} {\enquote {\bibinfo {title} {{Fermi-arc diversity on surface
  terminations of the magnetic Weyl semimetal Co$_3$Sn$_2$S$_2$}}}, }\href
  {\doibase 10.1126/science.aav2334} {\bibfield  {journal} {\bibinfo  {journal}
  {Science}\ }\textbf {\bibinfo {volume} {365}},\ \bibinfo {pages} {1286}
  (\bibinfo {year} {2019})}\BibitemShut {NoStop}%
\bibitem [{\citenamefont {Liu}\ \emph {et~al.}(2019)\citenamefont {Liu},
  \citenamefont {Liang}, \citenamefont {Liu}, \citenamefont {Xu}, \citenamefont
  {Li}, \citenamefont {Chen}, \citenamefont {Pei}, \citenamefont {Shi},
  \citenamefont {Mo}, \citenamefont {Dudin}, \citenamefont {Kim}, \citenamefont
  {Cacho}, \citenamefont {Li}, \citenamefont {Sun}, \citenamefont {Yang},
  \citenamefont {Liu}, \citenamefont {Parkin}, \citenamefont {Felser},\ and\
  \citenamefont {Chen}}]{Liu_magnetic_Weyl_2019}%
  \BibitemOpen
  \bibfield  {author} {\bibinfo {author} {\bibfnamefont {D.~F.}\ \bibnamefont
  {Liu}}, \bibinfo {author} {\bibfnamefont {A.~J.}\ \bibnamefont {Liang}},
  \bibinfo {author} {\bibfnamefont {E.~K.}\ \bibnamefont {Liu}}, \bibinfo
  {author} {\bibfnamefont {Q.~N.}\ \bibnamefont {Xu}}, \bibinfo {author}
  {\bibfnamefont {Y.~W.}\ \bibnamefont {Li}}, \bibinfo {author} {\bibfnamefont
  {C.}~\bibnamefont {Chen}},  \emph {et~al.},\ }\bibfield  {title} {\enquote
  {\bibinfo {title} {{Magnetic Weyl semimetal phase in a Kagom{\'e} crystal}}},
  }\href {\doibase 10.1126/science.aav2873} {\bibfield  {journal} {\bibinfo
  {journal} {Science}\ }\textbf {\bibinfo {volume} {365}},\ \bibinfo {pages}
  {1282} (\bibinfo {year} {2019})}\BibitemShut {NoStop}%
\bibitem [{\citenamefont {Benalcazar}\ \emph {et~al.}(2017)\citenamefont
  {Benalcazar}, \citenamefont {Bernevig},\ and\ \citenamefont
  {Hughes}}]{Benalcazar61}%
  \BibitemOpen
  \bibfield  {author} {\bibinfo {author} {\bibfnamefont {W.~A.}\ \bibnamefont
  {Benalcazar}}, \bibinfo {author} {\bibfnamefont {B.~A.}\ \bibnamefont
  {Bernevig}}, \ and\ \bibinfo {author} {\bibfnamefont {T.~L.}\ \bibnamefont
  {Hughes}},\ }\bibfield  {title} {\enquote {\bibinfo {title} {Quantized
  electric multipole insulators}}, }\href {\doibase 10.1126/science.aah6442}
  {\bibfield  {journal} {\bibinfo  {journal} {Science}\ }\textbf {\bibinfo
  {volume} {357}},\ \bibinfo {pages} {61} (\bibinfo {year} {2017})}\BibitemShut
  {NoStop}%
\bibitem [{\citenamefont {Song}\ \emph {et~al.}(2017)\citenamefont {Song},
  \citenamefont {Fang},\ and\ \citenamefont {Fang}}]{song_PRL_2017}%
  \BibitemOpen
  \bibfield  {author} {\bibinfo {author} {\bibfnamefont {Z.}~\bibnamefont
  {Song}}, \bibinfo {author} {\bibfnamefont {Z.}~\bibnamefont {Fang}}, \ and\
  \bibinfo {author} {\bibfnamefont {C.}~\bibnamefont {Fang}},\ }\bibfield
  {title} {\enquote {\bibinfo {title} {{$(d\ensuremath{-}2)$-Dimensional Edge
  States of Rotation Symmetry Protected Topological States}}}, }\href {\doibase
  10.1103/PhysRevLett.119.246402} {\bibfield  {journal} {\bibinfo  {journal}
  {Phys. Rev. Lett.}\ }\textbf {\bibinfo {volume} {119}},\ \bibinfo {pages}
  {246402} (\bibinfo {year} {2017})}\BibitemShut {NoStop}%
\bibitem [{\citenamefont {Langbehn}\ \emph {et~al.}(2017)\citenamefont
  {Langbehn}, \citenamefont {Peng}, \citenamefont {Trifunovic}, \citenamefont
  {von Oppen},\ and\ \citenamefont {Brouwer}}]{Piet_PRL_2017}%
  \BibitemOpen
  \bibfield  {author} {\bibinfo {author} {\bibfnamefont {J.}~\bibnamefont
  {Langbehn}}, \bibinfo {author} {\bibfnamefont {Y.}~\bibnamefont {Peng}},
  \bibinfo {author} {\bibfnamefont {L.}~\bibnamefont {Trifunovic}}, \bibinfo
  {author} {\bibfnamefont {F.}~\bibnamefont {von Oppen}}, \ and\ \bibinfo
  {author} {\bibfnamefont {P.~W.}\ \bibnamefont {Brouwer}},\ }\bibfield
  {title} {\enquote {\bibinfo {title} {{Reflection-Symmetric Second-Order
  Topological Insulators and Superconductors}}}, }\href {\doibase
  10.1103/PhysRevLett.119.246401} {\bibfield  {journal} {\bibinfo  {journal}
  {Phys. Rev. Lett.}\ }\textbf {\bibinfo {volume} {119}},\ \bibinfo {pages}
  {246401} (\bibinfo {year} {2017})}\BibitemShut {NoStop}%
\bibitem [{\citenamefont {Peterson}\ \emph {et~al.}(2020)\citenamefont
  {Peterson}, \citenamefont {Li}, \citenamefont {Benalcazar}, \citenamefont
  {Hughes},\ and\ \citenamefont {Bahl}}]{Peterson_fractional_2020}%
  \BibitemOpen
  \bibfield  {author} {\bibinfo {author} {\bibfnamefont {C.~W.}\ \bibnamefont
  {Peterson}}, \bibinfo {author} {\bibfnamefont {T.}~\bibnamefont {Li}},
  \bibinfo {author} {\bibfnamefont {W.~A.}\ \bibnamefont {Benalcazar}},
  \bibinfo {author} {\bibfnamefont {T.~L.}\ \bibnamefont {Hughes}}, \ and\
  \bibinfo {author} {\bibfnamefont {G.}~\bibnamefont {Bahl}},\ }\bibfield
  {title} {\enquote {\bibinfo {title} {A fractional corner anomaly reveals
  higher-order topology}}, }\href {\doibase 10.1126/science.aba7604} {\bibfield
   {journal} {\bibinfo  {journal} {Science}\ }\textbf {\bibinfo {volume}
  {368}},\ \bibinfo {pages} {1114} (\bibinfo {year} {2020})}\BibitemShut
  {NoStop}%
\bibitem [{\citenamefont {Schindler}\ \emph
  {et~al.}(2018{\natexlab{a}})\citenamefont {Schindler}, \citenamefont {Cook},
  \citenamefont {Vergniory}, \citenamefont {Wang}, \citenamefont {Parkin},
  \citenamefont {Bernevig},\ and\ \citenamefont {Neupert}}]{Schindl18science}%
  \BibitemOpen
  \bibfield  {author} {\bibinfo {author} {\bibfnamefont {F.}~\bibnamefont
  {Schindler}}, \bibinfo {author} {\bibfnamefont {A.~M.}\ \bibnamefont {Cook}},
  \bibinfo {author} {\bibfnamefont {M.~G.}\ \bibnamefont {Vergniory}}, \bibinfo
  {author} {\bibfnamefont {Z.}~\bibnamefont {Wang}}, \bibinfo {author}
  {\bibfnamefont {S.~S.~P.}\ \bibnamefont {Parkin}}, \bibinfo {author}
  {\bibfnamefont {B.~A.}\ \bibnamefont {Bernevig}}, \ and\ \bibinfo {author}
  {\bibfnamefont {T.}~\bibnamefont {Neupert}},\ }\bibfield  {title} {\enquote
  {\bibinfo {title} {Higher-order topological insulators}}, }\href {\doibase
  10.1126/sciadv.aat0346} {\bibfield  {journal} {\bibinfo  {journal} {Sci.
  Adv.}\ }\textbf {\bibinfo {volume} {4}} (\bibinfo {year}
  {2018}{\natexlab{a}}),\ 10.1126/sciadv.aat0346}\BibitemShut {NoStop}%
\bibitem [{\citenamefont {Schindler}\ \emph
  {et~al.}(2018{\natexlab{b}})\citenamefont {Schindler}, \citenamefont {Wang},
  \citenamefont {Vergniory}, \citenamefont {Cook}, \citenamefont {Murani},
  \citenamefont {Sengupta}, \citenamefont {Kasumov}, \citenamefont {Deblock},
  \citenamefont {Jeon}, \citenamefont {Drozdov} \emph
  {et~al.}}]{Schindler2018higher}%
  \BibitemOpen
  \bibfield  {author} {\bibinfo {author} {\bibfnamefont {F.}~\bibnamefont
  {Schindler}}, \bibinfo {author} {\bibfnamefont {Z.}~\bibnamefont {Wang}},
  \bibinfo {author} {\bibfnamefont {M.~G.}\ \bibnamefont {Vergniory}}, \bibinfo
  {author} {\bibfnamefont {A.~M.}\ \bibnamefont {Cook}}, \bibinfo {author}
  {\bibfnamefont {A.}~\bibnamefont {Murani}}, \bibinfo {author} {\bibfnamefont
  {S.}~\bibnamefont {Sengupta}},  \emph {et~al.},\ }\bibfield  {title}
  {\enquote {\bibinfo {title} {{Higher-order topology in bismuth}}}, }\href
  {https://www.nature.com/articles/s41567-018-0224-7} {\bibfield  {journal}
  {\bibinfo  {journal} {Nature physics}\ }\textbf {\bibinfo {volume} {14}},\
  \bibinfo {pages} {918} (\bibinfo {year} {2018}{\natexlab{b}})}\BibitemShut
  {NoStop}%
\bibitem [{\citenamefont {Serra-Garcia}\ \emph {et~al.}(2018)\citenamefont
  {Serra-Garcia}, \citenamefont {Peri}, \citenamefont {S{\"u}sstrunk},
  \citenamefont {Bilal}, \citenamefont {Larsen}, \citenamefont {Villanueva},\
  and\ \citenamefont {Huber}}]{serra2018observation}%
  \BibitemOpen
  \bibfield  {author} {\bibinfo {author} {\bibfnamefont {M.}~\bibnamefont
  {Serra-Garcia}}, \bibinfo {author} {\bibfnamefont {V.}~\bibnamefont {Peri}},
  \bibinfo {author} {\bibfnamefont {R.}~\bibnamefont {S{\"u}sstrunk}}, \bibinfo
  {author} {\bibfnamefont {O.~R.}\ \bibnamefont {Bilal}}, \bibinfo {author}
  {\bibfnamefont {T.}~\bibnamefont {Larsen}}, \bibinfo {author} {\bibfnamefont
  {L.~G.}\ \bibnamefont {Villanueva}}, \ and\ \bibinfo {author} {\bibfnamefont
  {S.~D.}\ \bibnamefont {Huber}},\ }\bibfield  {title} {\enquote {\bibinfo
  {title} {{Observation of a phononic quadrupole topological insulator}}},
  }\href {https://www.nature.com/articles/nature25156} {\bibfield  {journal}
  {\bibinfo  {journal} {Nature}\ }\textbf {\bibinfo {volume} {555}},\ \bibinfo
  {pages} {342} (\bibinfo {year} {2018})}\BibitemShut {NoStop}%
\bibitem [{\citenamefont {Ezawa}(2018)}]{Ezawa18PRL}%
  \BibitemOpen
  \bibfield  {author} {\bibinfo {author} {\bibfnamefont {M.}~\bibnamefont
  {Ezawa}},\ }\bibfield  {title} {\enquote {\bibinfo {title} {{Higher-Order
  Topological Insulators and Semimetals on the Breathing Kagome and Pyrochlore
  Lattices}}}, }\href {\doibase 10.1103/PhysRevLett.120.026801} {\bibfield
  {journal} {\bibinfo  {journal} {Phys. Rev. Lett.}\ }\textbf {\bibinfo
  {volume} {120}},\ \bibinfo {pages} {026801} (\bibinfo {year}
  {2018})}\BibitemShut {NoStop}%
\bibitem [{\citenamefont {Trifunovic}\ and\ \citenamefont
  {Brouwer}(2019)}]{Trifunovic19PRX}%
  \BibitemOpen
  \bibfield  {author} {\bibinfo {author} {\bibfnamefont {L.}~\bibnamefont
  {Trifunovic}}\ and\ \bibinfo {author} {\bibfnamefont {P.~W.}\ \bibnamefont
  {Brouwer}},\ }\bibfield  {title} {\enquote {\bibinfo {title} {{Higher-Order
  Bulk-Boundary Correspondence for Topological Crystalline Phases}}}, }\href
  {\doibase 10.1103/PhysRevX.9.011012} {\bibfield  {journal} {\bibinfo
  {journal} {Phys. Rev. X}\ }\textbf {\bibinfo {volume} {9}},\ \bibinfo {pages}
  {011012} (\bibinfo {year} {2019})}\BibitemShut {NoStop}%
\bibitem [{\citenamefont {Chen}\ \emph {et~al.}(2020)\citenamefont {Chen},
  \citenamefont {Chen}, \citenamefont {Gao}, \citenamefont {Zhou},\ and\
  \citenamefont {Xu}}]{ChenR20PRL}%
  \BibitemOpen
  \bibfield  {author} {\bibinfo {author} {\bibfnamefont {R.}~\bibnamefont
  {Chen}}, \bibinfo {author} {\bibfnamefont {C.-Z.}\ \bibnamefont {Chen}},
  \bibinfo {author} {\bibfnamefont {J.-H.}\ \bibnamefont {Gao}}, \bibinfo
  {author} {\bibfnamefont {B.}~\bibnamefont {Zhou}}, \ and\ \bibinfo {author}
  {\bibfnamefont {D.-H.}\ \bibnamefont {Xu}},\ }\bibfield  {title} {\enquote
  {\bibinfo {title} {{Higher-Order Topological Insulators in Quasicrystals}}},
  }\href {\doibase 10.1103/PhysRevLett.124.036803} {\bibfield  {journal}
  {\bibinfo  {journal} {Phys. Rev. Lett.}\ }\textbf {\bibinfo {volume} {124}},\
  \bibinfo {pages} {036803} (\bibinfo {year} {2020})}\BibitemShut {NoStop}%
\bibitem [{\citenamefont {Khalaf}(2018)}]{Eslam_PRB_2018}%
  \BibitemOpen
  \bibfield  {author} {\bibinfo {author} {\bibfnamefont {E.}~\bibnamefont
  {Khalaf}},\ }\bibfield  {title} {\enquote {\bibinfo {title} {Higher-order
  topological insulators and superconductors protected by inversion symmetry}},
  }\href {\doibase 10.1103/PhysRevB.97.205136} {\bibfield  {journal} {\bibinfo
  {journal} {Phys. Rev. B}\ }\textbf {\bibinfo {volume} {97}},\ \bibinfo
  {pages} {205136} (\bibinfo {year} {2018})}\BibitemShut {NoStop}%
\bibitem [{\citenamefont {Lin}\ and\ \citenamefont
  {Hughes}(2018)}]{Lin_PRB_2018}%
  \BibitemOpen
  \bibfield  {author} {\bibinfo {author} {\bibfnamefont {M.}~\bibnamefont
  {Lin}}\ and\ \bibinfo {author} {\bibfnamefont {T.~L.}\ \bibnamefont
  {Hughes}},\ }\bibfield  {title} {\enquote {\bibinfo {title} {Topological
  quadrupolar semimetals}}, }\href {\doibase 10.1103/PhysRevB.98.241103}
  {\bibfield  {journal} {\bibinfo  {journal} {Phys. Rev. B}\ }\textbf {\bibinfo
  {volume} {98}},\ \bibinfo {pages} {241103(R)} (\bibinfo {year}
  {2018})}\BibitemShut {NoStop}%
\bibitem [{\citenamefont {Zhang}\ \emph {et~al.}()\citenamefont {Zhang},
  \citenamefont {Hsu},\ and\ \citenamefont {Sarma}}]{zhang_2019_higherorder}%
  \BibitemOpen
  \bibfield  {author} {\bibinfo {author} {\bibfnamefont {R.-X.}\ \bibnamefont
  {Zhang}}, \bibinfo {author} {\bibfnamefont {Y.-T.}\ \bibnamefont {Hsu}}, \
  and\ \bibinfo {author} {\bibfnamefont {S.~D.}\ \bibnamefont {Sarma}},\
  }\bibfield  {title} {\enquote {\bibinfo {title} {Higher-order topological
  dirac superconductors}}, }\href@noop {} {\ }\Eprint
  {http://arxiv.org/abs/arXiv:1909.07980} {arXiv:1909.07980} \BibitemShut
  {NoStop}%
\bibitem [{\citenamefont {Wang}\ \emph {et~al.}()\citenamefont {Wang},
  \citenamefont {Lin}, \citenamefont {Jiang}, \citenamefont {Guo},\ and\
  \citenamefont {Jiang}}]{wang_2020_higherorder}%
  \BibitemOpen
  \bibfield  {author} {\bibinfo {author} {\bibfnamefont {H.-X.}\ \bibnamefont
  {Wang}}, \bibinfo {author} {\bibfnamefont {Z.-K.}\ \bibnamefont {Lin}},
  \bibinfo {author} {\bibfnamefont {B.}~\bibnamefont {Jiang}}, \bibinfo
  {author} {\bibfnamefont {G.-Y.}\ \bibnamefont {Guo}}, \ and\ \bibinfo
  {author} {\bibfnamefont {J.-H.}\ \bibnamefont {Jiang}},\ }\bibfield  {title}
  {\enquote {\bibinfo {title} {{Higher-Order Weyl Semimetals}}}, }\href@noop {}
  {\ }\Eprint {http://arxiv.org/abs/arXiv:2007.05068} {arXiv:2007.05068}
  \BibitemShut {NoStop}%
\bibitem [{\citenamefont {Ghorashi}\ \emph {et~al.}(2019)\citenamefont
  {Ghorashi}, \citenamefont {Hu}, \citenamefont {Hughes},\ and\ \citenamefont
  {Rossi}}]{ghorashi_second-order_2019}%
  \BibitemOpen
  \bibfield  {author} {\bibinfo {author} {\bibfnamefont {S.~A.~A.}\
  \bibnamefont {Ghorashi}}, \bibinfo {author} {\bibfnamefont {X.}~\bibnamefont
  {Hu}}, \bibinfo {author} {\bibfnamefont {T.~L.}\ \bibnamefont {Hughes}}, \
  and\ \bibinfo {author} {\bibfnamefont {E.}~\bibnamefont {Rossi}},\ }\bibfield
   {title} {\enquote {\bibinfo {title} {Second-order dirac superconductors and
  magnetic field induced majorana hinge modes}}, }\href {\doibase
  10.1103/PhysRevB.100.020509} {\bibfield  {journal} {\bibinfo  {journal}
  {Phys. Rev. B}\ }\textbf {\bibinfo {volume} {100}},\ \bibinfo {pages}
  {020509(R)} (\bibinfo {year} {2019})}\BibitemShut {NoStop}%
\bibitem [{\citenamefont {Ghorashi}\ \emph {et~al.}()\citenamefont {Ghorashi},
  \citenamefont {Li},\ and\ \citenamefont
  {Hughes}}]{ghorashi_2020_higherorder}%
  \BibitemOpen
  \bibfield  {author} {\bibinfo {author} {\bibfnamefont {S.~A.~A.}\
  \bibnamefont {Ghorashi}}, \bibinfo {author} {\bibfnamefont {T.}~\bibnamefont
  {Li}}, \ and\ \bibinfo {author} {\bibfnamefont {T.~L.}\ \bibnamefont
  {Hughes}},\ }\bibfield  {title} {\enquote {\bibinfo {title} {{Higher-order
  Weyl Semimetals}}}, }\href@noop {} {\ }\Eprint
  {http://arxiv.org/abs/arXiv:2007.02956} {arXiv:2007.02956} \BibitemShut
  {NoStop}%
\bibitem [{\citenamefont {Wang}\ \emph
  {et~al.}(2020{\natexlab{a}})\citenamefont {Wang}, \citenamefont {Dai},
  \citenamefont {Shao}, \citenamefont {Yang},\ and\ \citenamefont
  {Zhao}}]{wang2020boundary}%
  \BibitemOpen
  \bibfield  {author} {\bibinfo {author} {\bibfnamefont {K.}~\bibnamefont
  {Wang}}, \bibinfo {author} {\bibfnamefont {J.-X.}\ \bibnamefont {Dai}},
  \bibinfo {author} {\bibfnamefont {L.~B.}\ \bibnamefont {Shao}}, \bibinfo
  {author} {\bibfnamefont {S.~A.}\ \bibnamefont {Yang}}, \ and\ \bibinfo
  {author} {\bibfnamefont {Y.~X.}\ \bibnamefont {Zhao}},\ }\bibfield  {title}
  {\enquote {\bibinfo {title} {Boundary criticality of $\mathcal{PT}$-invariant
  topology and second-order nodal-line semimetals}}, }\href {\doibase
  10.1103/PhysRevLett.125.126403} {\bibfield  {journal} {\bibinfo  {journal}
  {Phys. Rev. Lett.}\ }\textbf {\bibinfo {volume} {125}},\ \bibinfo {pages}
  {126403} (\bibinfo {year} {2020}{\natexlab{a}})}\BibitemShut {NoStop}%
\bibitem [{\citenamefont {Cho}\ \emph {et~al.}(2012)\citenamefont {Cho},
  \citenamefont {Bardarson}, \citenamefont {Lu},\ and\ \citenamefont
  {Moore}}]{Cho12PRB}%
  \BibitemOpen
  \bibfield  {author} {\bibinfo {author} {\bibfnamefont {G.~Y.}\ \bibnamefont
  {Cho}}, \bibinfo {author} {\bibfnamefont {J.~H.}\ \bibnamefont {Bardarson}},
  \bibinfo {author} {\bibfnamefont {Y.-M.}\ \bibnamefont {Lu}}, \ and\ \bibinfo
  {author} {\bibfnamefont {J.~E.}\ \bibnamefont {Moore}},\ }\bibfield  {title}
  {\enquote {\bibinfo {title} {{Superconductivity of doped Weyl semimetals:
  Finite-momentum pairing and electronic analog of the ${}^{3}$He-$A$ phase}}},
  }\href {\doibase 10.1103/PhysRevB.86.214514} {\bibfield  {journal} {\bibinfo
  {journal} {Phys. Rev. B}\ }\textbf {\bibinfo {volume} {86}},\ \bibinfo
  {pages} {214514} (\bibinfo {year} {2012})}\BibitemShut {NoStop}%
\bibitem [{\citenamefont {Bednik}\ \emph {et~al.}(2015)\citenamefont {Bednik},
  \citenamefont {Zyuzin},\ and\ \citenamefont {Burkov}}]{Bednik15PRB}%
  \BibitemOpen
  \bibfield  {author} {\bibinfo {author} {\bibfnamefont {G.}~\bibnamefont
  {Bednik}}, \bibinfo {author} {\bibfnamefont {A.~A.}\ \bibnamefont {Zyuzin}},
  \ and\ \bibinfo {author} {\bibfnamefont {A.~A.}\ \bibnamefont {Burkov}},\
  }\bibfield  {title} {\enquote {\bibinfo {title} {{Superconductivity in Weyl
  metals}}}, }\href {\doibase 10.1103/PhysRevB.92.035153} {\bibfield  {journal}
  {\bibinfo  {journal} {Phys. Rev. B}\ }\textbf {\bibinfo {volume} {92}},\
  \bibinfo {pages} {035153} (\bibinfo {year} {2015})}\BibitemShut {NoStop}%
\bibitem [{\citenamefont {Biswas}\ \emph {et~al.}(2013)\citenamefont {Biswas},
  \citenamefont {Luetkens}, \citenamefont {Neupert}, \citenamefont {St\"urzer},
  \citenamefont {Baines}, \citenamefont {Pascua}, \citenamefont {Schnyder},
  \citenamefont {Fischer}, \citenamefont {Goryo}, \citenamefont {Lees},
  \citenamefont {Maeter}, \citenamefont {Br\"uckner}, \citenamefont {Klauss},
  \citenamefont {Nicklas}, \citenamefont {Baker}, \citenamefont {Hillier},
  \citenamefont {Sigrist}, \citenamefont {Amato},\ and\ \citenamefont
  {Johrendt}}]{Biswas_PRB_2013}%
  \BibitemOpen
  \bibfield  {author} {\bibinfo {author} {\bibfnamefont {P.~K.}\ \bibnamefont
  {Biswas}}, \bibinfo {author} {\bibfnamefont {H.}~\bibnamefont {Luetkens}},
  \bibinfo {author} {\bibfnamefont {T.}~\bibnamefont {Neupert}}, \bibinfo
  {author} {\bibfnamefont {T.}~\bibnamefont {St\"urzer}}, \bibinfo {author}
  {\bibfnamefont {C.}~\bibnamefont {Baines}}, \bibinfo {author} {\bibfnamefont
  {G.}~\bibnamefont {Pascua}},  \emph {et~al.},\ }\bibfield  {title} {\enquote
  {\bibinfo {title} {{Evidence for superconductivity with broken time-reversal
  symmetry in locally noncentrosymmetric SrPtAs}}}, }\href {\doibase
  10.1103/PhysRevB.87.180503} {\bibfield  {journal} {\bibinfo  {journal} {Phys.
  Rev. B}\ }\textbf {\bibinfo {volume} {87}},\ \bibinfo {pages} {180503(R)}
  (\bibinfo {year} {2013})}\BibitemShut {NoStop}%
\bibitem [{\citenamefont {Fischer}\ \emph {et~al.}(2014)\citenamefont
  {Fischer}, \citenamefont {Neupert}, \citenamefont {Platt}, \citenamefont
  {Schnyder}, \citenamefont {Hanke}, \citenamefont {Goryo}, \citenamefont
  {Thomale},\ and\ \citenamefont {Sigrist}}]{Fischer_PRB_2014}%
  \BibitemOpen
  \bibfield  {author} {\bibinfo {author} {\bibfnamefont {M.~H.}\ \bibnamefont
  {Fischer}}, \bibinfo {author} {\bibfnamefont {T.}~\bibnamefont {Neupert}},
  \bibinfo {author} {\bibfnamefont {C.}~\bibnamefont {Platt}}, \bibinfo
  {author} {\bibfnamefont {A.~P.}\ \bibnamefont {Schnyder}}, \bibinfo {author}
  {\bibfnamefont {W.}~\bibnamefont {Hanke}}, \bibinfo {author} {\bibfnamefont
  {J.}~\bibnamefont {Goryo}}, \bibinfo {author} {\bibfnamefont
  {R.}~\bibnamefont {Thomale}}, \ and\ \bibinfo {author} {\bibfnamefont
  {M.}~\bibnamefont {Sigrist}},\ }\bibfield  {title} {\enquote {\bibinfo
  {title} {{Chiral $d$-wave superconductivity in SrPtAs}}}, }\href {\doibase
  10.1103/PhysRevB.89.020509} {\bibfield  {journal} {\bibinfo  {journal} {Phys.
  Rev. B}\ }\textbf {\bibinfo {volume} {89}},\ \bibinfo {pages} {020509(R)}
  (\bibinfo {year} {2014})}\BibitemShut {NoStop}%
\bibitem [{\citenamefont {Goswami}\ and\ \citenamefont
  {Balicas}()}]{goswami_2013_topological}%
  \BibitemOpen
  \bibfield  {author} {\bibinfo {author} {\bibfnamefont {P.}~\bibnamefont
  {Goswami}}\ and\ \bibinfo {author} {\bibfnamefont {L.}~\bibnamefont
  {Balicas}},\ }\bibfield  {title} {\enquote {\bibinfo {title} {{Topological
  properties of possible Weyl superconducting states of
  URu$_\mathbf{2}$Si$_\mathbf{2}$}}}, }\href@noop {} {\ }\Eprint
  {http://arxiv.org/abs/arXiv:1312.3632} {arXiv:1312.3632} \BibitemShut
  {NoStop}%
\bibitem [{\citenamefont {Hayes}\ \emph {et~al.}()\citenamefont {Hayes},
  \citenamefont {Wei}, \citenamefont {Metz}, \citenamefont {Zhang},
  \citenamefont {Eo}, \citenamefont {Ran}, \citenamefont {Saha}, \citenamefont
  {Collini}, \citenamefont {Butch}, \citenamefont {Agterberg}, \citenamefont
  {Kapitulnik},\ and\ \citenamefont {Paglione}}]{hayes_2020_weyl}%
  \BibitemOpen
  \bibfield  {author} {\bibinfo {author} {\bibfnamefont {I.~M.}\ \bibnamefont
  {Hayes}}, \bibinfo {author} {\bibfnamefont {D.~S.}\ \bibnamefont {Wei}},
  \bibinfo {author} {\bibfnamefont {T.}~\bibnamefont {Metz}}, \bibinfo {author}
  {\bibfnamefont {J.}~\bibnamefont {Zhang}}, \bibinfo {author} {\bibfnamefont
  {Y.~S.}\ \bibnamefont {Eo}}, \bibinfo {author} {\bibfnamefont
  {S.}~\bibnamefont {Ran}},  \emph {et~al.},\ }\bibfield  {title} {\enquote
  {\bibinfo {title} {{Weyl Superconductivity in UTe$_2$}}}, }\href@noop {} {\
  }\Eprint {http://arxiv.org/abs/arXiv:2002.02539} {arXiv:2002.02539}
  \BibitemShut {NoStop}%
\bibitem [{\citenamefont {Wang}\ \emph
  {et~al.}(2020{\natexlab{b}})\citenamefont {Wang}, \citenamefont {Kim},
  \citenamefont {Liu}, \citenamefont {Cevallos}, \citenamefont {Cava},\ and\
  \citenamefont {Ong}}]{wang_evidence_2020}%
  \BibitemOpen
  \bibfield  {author} {\bibinfo {author} {\bibfnamefont {W.}~\bibnamefont
  {Wang}}, \bibinfo {author} {\bibfnamefont {S.}~\bibnamefont {Kim}}, \bibinfo
  {author} {\bibfnamefont {M.}~\bibnamefont {Liu}}, \bibinfo {author}
  {\bibfnamefont {F.~A.}\ \bibnamefont {Cevallos}}, \bibinfo {author}
  {\bibfnamefont {R.~J.}\ \bibnamefont {Cava}}, \ and\ \bibinfo {author}
  {\bibfnamefont {N.~P.}\ \bibnamefont {Ong}},\ }\bibfield  {title} {\enquote
  {\bibinfo {title} {Evidence for an edge supercurrent in the weyl
  superconductor {MoTe}2}}, }\href {\doibase 10.1126/science.aaw9270}
  {\bibfield  {journal} {\bibinfo  {journal} {Science}\ }\textbf {\bibinfo
  {volume} {368}},\ \bibinfo {pages} {534} (\bibinfo {year}
  {2020}{\natexlab{b}})}\BibitemShut {NoStop}%
\bibitem [{\citenamefont {Yamashita}\ \emph {et~al.}(2015)\citenamefont
  {Yamashita}, \citenamefont {Shimoyama}, \citenamefont {Haga}, \citenamefont
  {Matsuda}, \citenamefont {Yamamoto}, \citenamefont {Onuki}, \citenamefont
  {Sumiyoshi}, \citenamefont {Fujimoto}, \citenamefont {Levchenko},
  \citenamefont {Shibauchi},\ and\ \citenamefont
  {Matsuda}}]{yamashita_colossal_2015}%
  \BibitemOpen
  \bibfield  {author} {\bibinfo {author} {\bibfnamefont {T.}~\bibnamefont
  {Yamashita}}, \bibinfo {author} {\bibfnamefont {Y.}~\bibnamefont
  {Shimoyama}}, \bibinfo {author} {\bibfnamefont {Y.}~\bibnamefont {Haga}},
  \bibinfo {author} {\bibfnamefont {T.~D.}\ \bibnamefont {Matsuda}}, \bibinfo
  {author} {\bibfnamefont {E.}~\bibnamefont {Yamamoto}}, \bibinfo {author}
  {\bibfnamefont {Y.}~\bibnamefont {Onuki}},  \emph {et~al.},\ }\bibfield
  {title} {\enquote {\bibinfo {title} {{Colossal thermomagnetic response in the
  exotic superconductor {URu}2Si2}}}, }\href {\doibase 10.1038/nphys3170}
  {\bibfield  {journal} {\bibinfo  {journal} {Nature Physics}\ }\textbf
  {\bibinfo {volume} {11}},\ \bibinfo {pages} {17} (\bibinfo {year}
  {2015})}\BibitemShut {NoStop}%
\bibitem [{\citenamefont {Yanase}\ and\ \citenamefont
  {Shiozaki}(2017)}]{Yanase_2017_PRB}%
  \BibitemOpen
  \bibfield  {author} {\bibinfo {author} {\bibfnamefont {Y.}~\bibnamefont
  {Yanase}}\ and\ \bibinfo {author} {\bibfnamefont {K.}~\bibnamefont
  {Shiozaki}},\ }\bibfield  {title} {\enquote {\bibinfo {title} {{M\"obius
  topological superconductivity in UPt$_{3}$}}}, }\href {\doibase
  10.1103/PhysRevB.95.224514} {\bibfield  {journal} {\bibinfo  {journal} {Phys.
  Rev. B}\ }\textbf {\bibinfo {volume} {95}},\ \bibinfo {pages} {224514}
  (\bibinfo {year} {2017})}\BibitemShut {NoStop}%
\bibitem [{\citenamefont {Yanase}(2016)}]{Yanase_PRB_2016}%
  \BibitemOpen
  \bibfield  {author} {\bibinfo {author} {\bibfnamefont {Y.}~\bibnamefont
  {Yanase}},\ }\bibfield  {title} {\enquote {\bibinfo {title} {{Nonsymmorphic
  Weyl superconductivity in ${\mathrm{UPt}}_{3}$ based on ${E}_{2u}$
  representation}}}, }\href {\doibase 10.1103/PhysRevB.94.174502} {\bibfield
  {journal} {\bibinfo  {journal} {Phys. Rev. B}\ }\textbf {\bibinfo {volume}
  {94}},\ \bibinfo {pages} {174502} (\bibinfo {year} {2016})}\BibitemShut
  {NoStop}%
\bibitem [{\citenamefont {Yuan}\ \emph {et~al.}(2017)\citenamefont {Yuan},
  \citenamefont {He},\ and\ \citenamefont {Law}}]{Yuan_PRB_2017}%
  \BibitemOpen
  \bibfield  {author} {\bibinfo {author} {\bibfnamefont {N.~F.~Q.}\
  \bibnamefont {Yuan}}, \bibinfo {author} {\bibfnamefont {W.-Y.}\ \bibnamefont
  {He}}, \ and\ \bibinfo {author} {\bibfnamefont {K.~T.}\ \bibnamefont {Law}},\
  }\bibfield  {title} {\enquote {\bibinfo {title} {{Superconductivity-induced
  ferromagnetism and Weyl superconductivity in Nb-doped Bi$_{2}$Se$_{3}$}}},
  }\href {\doibase 10.1103/PhysRevB.95.201109} {\bibfield  {journal} {\bibinfo
  {journal} {Phys. Rev. B}\ }\textbf {\bibinfo {volume} {95}},\ \bibinfo
  {pages} {201109(R)} (\bibinfo {year} {2017})}\BibitemShut {NoStop}%
\bibitem [{\citenamefont {Hsu}\ \emph {et~al.}(2018)\citenamefont {Hsu},
  \citenamefont {Stano}, \citenamefont {Klinovaja},\ and\ \citenamefont
  {Loss}}]{CHHsu18PRL}%
  \BibitemOpen
  \bibfield  {author} {\bibinfo {author} {\bibfnamefont {C.-H.}\ \bibnamefont
  {Hsu}}, \bibinfo {author} {\bibfnamefont {P.}~\bibnamefont {Stano}}, \bibinfo
  {author} {\bibfnamefont {J.}~\bibnamefont {Klinovaja}}, \ and\ \bibinfo
  {author} {\bibfnamefont {D.}~\bibnamefont {Loss}},\ }\bibfield  {title}
  {\enquote {\bibinfo {title} {{Majorana Kramers Pairs in Higher-Order
  Topological Insulators}}}, }\href {\doibase 10.1103/PhysRevLett.121.196801}
  {\bibfield  {journal} {\bibinfo  {journal} {Phys. Rev. Lett.}\ }\textbf
  {\bibinfo {volume} {121}},\ \bibinfo {pages} {196801} (\bibinfo {year}
  {2018})}\BibitemShut {NoStop}%
\bibitem [{\citenamefont {Liu}\ \emph {et~al.}(2018)\citenamefont {Liu},
  \citenamefont {He},\ and\ \citenamefont {Nori}}]{LiuT18PRB}%
  \BibitemOpen
  \bibfield  {author} {\bibinfo {author} {\bibfnamefont {T.}~\bibnamefont
  {Liu}}, \bibinfo {author} {\bibfnamefont {J.~J.}\ \bibnamefont {He}}, \ and\
  \bibinfo {author} {\bibfnamefont {F.}~\bibnamefont {Nori}},\ }\bibfield
  {title} {\enquote {\bibinfo {title} {Majorana corner states in a
  two-dimensional magnetic topological insulator on a high-temperature
  superconductor}}, }\href {\doibase 10.1103/PhysRevB.98.245413} {\bibfield
  {journal} {\bibinfo  {journal} {Phys. Rev. B}\ }\textbf {\bibinfo {volume}
  {98}},\ \bibinfo {pages} {245413} (\bibinfo {year} {2018})}\BibitemShut
  {NoStop}%
\bibitem [{\citenamefont {Yan}\ \emph {et~al.}(2018)\citenamefont {Yan},
  \citenamefont {Song},\ and\ \citenamefont {Wang}}]{YanZB18PRL}%
  \BibitemOpen
  \bibfield  {author} {\bibinfo {author} {\bibfnamefont {Z.}~\bibnamefont
  {Yan}}, \bibinfo {author} {\bibfnamefont {F.}~\bibnamefont {Song}}, \ and\
  \bibinfo {author} {\bibfnamefont {Z.}~\bibnamefont {Wang}},\ }\bibfield
  {title} {\enquote {\bibinfo {title} {{Majorana Corner Modes in a
  High-Temperature Platform}}}, }\href {\doibase
  10.1103/PhysRevLett.121.096803} {\bibfield  {journal} {\bibinfo  {journal}
  {Phys. Rev. Lett.}\ }\textbf {\bibinfo {volume} {121}},\ \bibinfo {pages}
  {096803} (\bibinfo {year} {2018})}\BibitemShut {NoStop}%
\bibitem [{\citenamefont {Wang}\ \emph {et~al.}(2018)\citenamefont {Wang},
  \citenamefont {Liu}, \citenamefont {Lu},\ and\ \citenamefont
  {Zhang}}]{WangQY18PRL}%
  \BibitemOpen
  \bibfield  {author} {\bibinfo {author} {\bibfnamefont {Q.}~\bibnamefont
  {Wang}}, \bibinfo {author} {\bibfnamefont {C.-C.}\ \bibnamefont {Liu}},
  \bibinfo {author} {\bibfnamefont {Y.-M.}\ \bibnamefont {Lu}}, \ and\ \bibinfo
  {author} {\bibfnamefont {F.}~\bibnamefont {Zhang}},\ }\bibfield  {title}
  {\enquote {\bibinfo {title} {{High-Temperature Majorana Corner States}}},
  }\href {\doibase 10.1103/PhysRevLett.121.186801} {\bibfield  {journal}
  {\bibinfo  {journal} {Phys. Rev. Lett.}\ }\textbf {\bibinfo {volume} {121}},\
  \bibinfo {pages} {186801} (\bibinfo {year} {2018})}\BibitemShut {NoStop}%
\bibitem [{\citenamefont {Zhu}(2018)}]{ZhuXY18PRB}%
  \BibitemOpen
  \bibfield  {author} {\bibinfo {author} {\bibfnamefont {X.}~\bibnamefont
  {Zhu}},\ }\bibfield  {title} {\enquote {\bibinfo {title} {{Tunable Majorana
  corner states in a two-dimensional second-order topological superconductor
  induced by magnetic fields}}}, }\href {\doibase 10.1103/PhysRevB.97.205134}
  {\bibfield  {journal} {\bibinfo  {journal} {Phys. Rev. B}\ }\textbf {\bibinfo
  {volume} {97}},\ \bibinfo {pages} {205134} (\bibinfo {year}
  {2018})}\BibitemShut {NoStop}%
\bibitem [{\citenamefont {Zhu}(2019)}]{zhu_PRL_2019}%
  \BibitemOpen
  \bibfield  {author} {\bibinfo {author} {\bibfnamefont {X.}~\bibnamefont
  {Zhu}},\ }\bibfield  {title} {\enquote {\bibinfo {title} {{Second-Order
  Topological Superconductors with Mixed Pairing}}}, }\href {\doibase
  10.1103/PhysRevLett.122.236401} {\bibfield  {journal} {\bibinfo  {journal}
  {Phys. Rev. Lett.}\ }\textbf {\bibinfo {volume} {122}},\ \bibinfo {pages}
  {236401} (\bibinfo {year} {2019})}\BibitemShut {NoStop}%
\bibitem [{\citenamefont {Geier}\ \emph {et~al.}(2018)\citenamefont {Geier},
  \citenamefont {Trifunovic}, \citenamefont {Hoskam},\ and\ \citenamefont
  {Brouwer}}]{Geier18PRB}%
  \BibitemOpen
  \bibfield  {author} {\bibinfo {author} {\bibfnamefont {M.}~\bibnamefont
  {Geier}}, \bibinfo {author} {\bibfnamefont {L.}~\bibnamefont {Trifunovic}},
  \bibinfo {author} {\bibfnamefont {M.}~\bibnamefont {Hoskam}}, \ and\ \bibinfo
  {author} {\bibfnamefont {P.~W.}\ \bibnamefont {Brouwer}},\ }\bibfield
  {title} {\enquote {\bibinfo {title} {{Second-order topological insulators and
  superconductors with an order-two crystalline symmetry}}}, }\href {\doibase
  10.1103/PhysRevB.97.205135} {\bibfield  {journal} {\bibinfo  {journal} {Phys.
  Rev. B}\ }\textbf {\bibinfo {volume} {97}},\ \bibinfo {pages} {205135}
  (\bibinfo {year} {2018})}\BibitemShut {NoStop}%
\bibitem [{\citenamefont {Yan}(2019)}]{yan_PRL_2019}%
  \BibitemOpen
  \bibfield  {author} {\bibinfo {author} {\bibfnamefont {Z.}~\bibnamefont
  {Yan}},\ }\bibfield  {title} {\enquote {\bibinfo {title} {{Higher-Order
  Topological Odd-Parity Superconductors}}}, }\href {\doibase
  10.1103/PhysRevLett.123.177001} {\bibfield  {journal} {\bibinfo  {journal}
  {Phys. Rev. Lett.}\ }\textbf {\bibinfo {volume} {123}},\ \bibinfo {pages}
  {177001} (\bibinfo {year} {2019})}\BibitemShut {NoStop}%
\bibitem [{\citenamefont {Volpez}\ \emph {et~al.}(2019)\citenamefont {Volpez},
  \citenamefont {Loss},\ and\ \citenamefont {Klinovaja}}]{Volpez19PRL}%
  \BibitemOpen
  \bibfield  {author} {\bibinfo {author} {\bibfnamefont {Y.}~\bibnamefont
  {Volpez}}, \bibinfo {author} {\bibfnamefont {D.}~\bibnamefont {Loss}}, \ and\
  \bibinfo {author} {\bibfnamefont {J.}~\bibnamefont {Klinovaja}},\ }\bibfield
  {title} {\enquote {\bibinfo {title} {{Second-Order Topological
  Superconductivity in $\ensuremath{\pi}$-Junction Rashba Layers}}}, }\href
  {\doibase 10.1103/PhysRevLett.122.126402} {\bibfield  {journal} {\bibinfo
  {journal} {Phys. Rev. Lett.}\ }\textbf {\bibinfo {volume} {122}},\ \bibinfo
  {pages} {126402} (\bibinfo {year} {2019})}\BibitemShut {NoStop}%
\bibitem [{\citenamefont {Zhang}\ \emph {et~al.}(2019)\citenamefont {Zhang},
  \citenamefont {Cole},\ and\ \citenamefont {Das~Sarma}}]{RXZhang19PRL}%
  \BibitemOpen
  \bibfield  {author} {\bibinfo {author} {\bibfnamefont {R.-X.}\ \bibnamefont
  {Zhang}}, \bibinfo {author} {\bibfnamefont {W.~S.}\ \bibnamefont {Cole}}, \
  and\ \bibinfo {author} {\bibfnamefont {S.}~\bibnamefont {Das~Sarma}},\
  }\bibfield  {title} {\enquote {\bibinfo {title} {{Helical Hinge Majorana
  Modes in Iron-Based Superconductors}}}, }\href {\doibase
  10.1103/PhysRevLett.122.187001} {\bibfield  {journal} {\bibinfo  {journal}
  {Phys. Rev. Lett.}\ }\textbf {\bibinfo {volume} {122}},\ \bibinfo {pages}
  {187001} (\bibinfo {year} {2019})}\BibitemShut {NoStop}%
\bibitem [{\citenamefont {Zhang}\ and\ \citenamefont
  {Trauzettel}(2020)}]{ZhangSB20PRR}%
  \BibitemOpen
  \bibfield  {author} {\bibinfo {author} {\bibfnamefont {S.-B.}\ \bibnamefont
  {Zhang}}\ and\ \bibinfo {author} {\bibfnamefont {B.}~\bibnamefont
  {Trauzettel}},\ }\bibfield  {title} {\enquote {\bibinfo {title} {{Detection
  of second-order topological superconductors by Josephson junctions}}}, }\href
  {\doibase 10.1103/PhysRevResearch.2.012018} {\bibfield  {journal} {\bibinfo
  {journal} {Phys. Rev. Research}\ }\textbf {\bibinfo {volume} {2}},\ \bibinfo
  {pages} {012018} (\bibinfo {year} {2020})}\BibitemShut {NoStop}%
\bibitem [{\citenamefont {Plekhanov}\ \emph {et~al.}(2019)\citenamefont
  {Plekhanov}, \citenamefont {Thakurathi}, \citenamefont {Loss},\ and\
  \citenamefont {Klinovaja}}]{Plekhanov19PRR}%
  \BibitemOpen
  \bibfield  {author} {\bibinfo {author} {\bibfnamefont {K.}~\bibnamefont
  {Plekhanov}}, \bibinfo {author} {\bibfnamefont {M.}~\bibnamefont
  {Thakurathi}}, \bibinfo {author} {\bibfnamefont {D.}~\bibnamefont {Loss}}, \
  and\ \bibinfo {author} {\bibfnamefont {J.}~\bibnamefont {Klinovaja}},\
  }\bibfield  {title} {\enquote {\bibinfo {title} {{Floquet second-order
  topological superconductor driven via ferromagnetic resonance}}}, }\href
  {\doibase 10.1103/PhysRevResearch.1.032013} {\bibfield  {journal} {\bibinfo
  {journal} {Phys. Rev. Research}\ }\textbf {\bibinfo {volume} {1}},\ \bibinfo
  {pages} {032013} (\bibinfo {year} {2019})}\BibitemShut {NoStop}%
\bibitem [{\citenamefont {Wu}\ \emph {et~al.}(2020)\citenamefont {Wu},
  \citenamefont {Hou}, \citenamefont {Li}, \citenamefont {Luo}, \citenamefont
  {Shi},\ and\ \citenamefont {Zhang}}]{WUYJ20PRL}%
  \BibitemOpen
  \bibfield  {author} {\bibinfo {author} {\bibfnamefont {Y.-J.}\ \bibnamefont
  {Wu}}, \bibinfo {author} {\bibfnamefont {J.}~\bibnamefont {Hou}}, \bibinfo
  {author} {\bibfnamefont {Y.-M.}\ \bibnamefont {Li}}, \bibinfo {author}
  {\bibfnamefont {X.-W.}\ \bibnamefont {Luo}}, \bibinfo {author} {\bibfnamefont
  {X.}~\bibnamefont {Shi}}, \ and\ \bibinfo {author} {\bibfnamefont
  {C.}~\bibnamefont {Zhang}},\ }\bibfield  {title} {\enquote {\bibinfo {title}
  {{In-Plane Zeeman-Field-Induced Majorana Corner and Hinge Modes in an
  $s$-Wave Superconductor Heterostructure}}}, }\href {\doibase
  10.1103/PhysRevLett.124.227001} {\bibfield  {journal} {\bibinfo  {journal}
  {Phys. Rev. Lett.}\ }\textbf {\bibinfo {volume} {124}},\ \bibinfo {pages}
  {227001} (\bibinfo {year} {2020})}\BibitemShut {NoStop}%
\bibitem [{\citenamefont {Pan}\ \emph {et~al.}(2019)\citenamefont {Pan},
  \citenamefont {Yang}, \citenamefont {Chen}, \citenamefont {Xu}, \citenamefont
  {Liu},\ and\ \citenamefont {Liu}}]{PanXH19PRL}%
  \BibitemOpen
  \bibfield  {author} {\bibinfo {author} {\bibfnamefont {X.-H.}\ \bibnamefont
  {Pan}}, \bibinfo {author} {\bibfnamefont {K.-J.}\ \bibnamefont {Yang}},
  \bibinfo {author} {\bibfnamefont {L.}~\bibnamefont {Chen}}, \bibinfo {author}
  {\bibfnamefont {G.}~\bibnamefont {Xu}}, \bibinfo {author} {\bibfnamefont
  {C.-X.}\ \bibnamefont {Liu}}, \ and\ \bibinfo {author} {\bibfnamefont
  {X.}~\bibnamefont {Liu}},\ }\bibfield  {title} {\enquote {\bibinfo {title}
  {{Lattice-Symmetry-Assisted Second-Order Topological Superconductors and
  Majorana Patterns}}}, }\href {\doibase 10.1103/PhysRevLett.123.156801}
  {\bibfield  {journal} {\bibinfo  {journal} {Phys. Rev. Lett.}\ }\textbf
  {\bibinfo {volume} {123}},\ \bibinfo {pages} {156801} (\bibinfo {year}
  {2019})}\BibitemShut {NoStop}%
\bibitem [{\citenamefont {Ahn}\ and\ \citenamefont
  {Yang}(2020)}]{Yang_PRR_2020}%
  \BibitemOpen
  \bibfield  {author} {\bibinfo {author} {\bibfnamefont {J.}~\bibnamefont
  {Ahn}}\ and\ \bibinfo {author} {\bibfnamefont {B.-J.}\ \bibnamefont {Yang}},\
  }\bibfield  {title} {\enquote {\bibinfo {title} {Higher-order topological
  superconductivity of spin-polarized fermions}}, }\href {\doibase
  10.1103/PhysRevResearch.2.012060} {\bibfield  {journal} {\bibinfo  {journal}
  {Phys. Rev. Research}\ }\textbf {\bibinfo {volume} {2}},\ \bibinfo {pages}
  {012060} (\bibinfo {year} {2020})}\BibitemShut {NoStop}%
\bibitem [{\citenamefont {{Zhang}}\ \emph {et~al.}(2020)\citenamefont
  {{Zhang}}, \citenamefont {{Calzona}},\ and\ \citenamefont
  {{Trauzettel}}}]{ZhangSB20arxiv2}%
  \BibitemOpen
  \bibfield  {author} {\bibinfo {author} {\bibfnamefont {S.-B.}\ \bibnamefont
  {{Zhang}}}, \bibinfo {author} {\bibfnamefont {A.}~\bibnamefont {{Calzona}}},
  \ and\ \bibinfo {author} {\bibfnamefont {B.}~\bibnamefont {{Trauzettel}}},\
  }\bibfield  {title} {\enquote {\bibinfo {title} {{All-electrically tunable
  networks of Majorana bound states}}}, }\href {\doibase
  10.1103/PhysRevB.102.100503} {\bibfield  {journal} {\bibinfo  {journal}
  {Phys. Rev. B}\ }\textbf {\bibinfo {volume} {102}},\ \bibinfo {pages}
  {100503(R)} (\bibinfo {year} {2020})}\BibitemShut {NoStop}%
\bibitem [{\citenamefont {{Zhang}}\ \emph {et~al.}()\citenamefont {{Zhang}},
  \citenamefont {{Rui}}, \citenamefont {{Calzona}}, \citenamefont {{Choi}},
  \citenamefont {{Schnyder}},\ and\ \citenamefont
  {{Trauzettel}}}]{ZhangSB20arxiv1}%
  \BibitemOpen
  \bibfield  {author} {\bibinfo {author} {\bibfnamefont {S.-B.}\ \bibnamefont
  {{Zhang}}}, \bibinfo {author} {\bibfnamefont {W.~B.}\ \bibnamefont {{Rui}}},
  \bibinfo {author} {\bibfnamefont {A.}~\bibnamefont {{Calzona}}}, \bibinfo
  {author} {\bibfnamefont {S.-J.}\ \bibnamefont {{Choi}}}, \bibinfo {author}
  {\bibfnamefont {A.~P.}\ \bibnamefont {{Schnyder}}}, \ and\ \bibinfo {author}
  {\bibfnamefont {B.}~\bibnamefont {{Trauzettel}}},\ }\bibfield  {title}
  {\enquote {\bibinfo {title} {{Topological and holonomic quantum computation
  based on second-order topological superconductors}}}, }\href@noop {} {\
  }\Eprint {http://arxiv.org/abs/2002.05741} {arXiv:2002.05741} \BibitemShut
  {NoStop}%
\bibitem [{\citenamefont {Tiwari}\ \emph {et~al.}(2020)\citenamefont {Tiwari},
  \citenamefont {Li}, \citenamefont {Bernevig}, \citenamefont {Neupert},\ and\
  \citenamefont {Parameswaran}}]{Tiwari20PRL}%
  \BibitemOpen
  \bibfield  {author} {\bibinfo {author} {\bibfnamefont {A.}~\bibnamefont
  {Tiwari}}, \bibinfo {author} {\bibfnamefont {M.-H.}\ \bibnamefont {Li}},
  \bibinfo {author} {\bibfnamefont {B.~A.}\ \bibnamefont {Bernevig}}, \bibinfo
  {author} {\bibfnamefont {T.}~\bibnamefont {Neupert}}, \ and\ \bibinfo
  {author} {\bibfnamefont {S.~A.}\ \bibnamefont {Parameswaran}},\ }\bibfield
  {title} {\enquote {\bibinfo {title} {{Unhinging the Surfaces of Higher-Order
  Topological Insulators and Superconductors}}}, }\href {\doibase
  10.1103/PhysRevLett.124.046801} {\bibfield  {journal} {\bibinfo  {journal}
  {Phys. Rev. Lett.}\ }\textbf {\bibinfo {volume} {124}},\ \bibinfo {pages}
  {046801} (\bibinfo {year} {2020})}\BibitemShut {NoStop}%
\bibitem [{\citenamefont {Laubscher}\ \emph {et~al.}(2020)\citenamefont
  {Laubscher}, \citenamefont {Loss},\ and\ \citenamefont
  {Klinovaja}}]{Laubscher20PRR}%
  \BibitemOpen
  \bibfield  {author} {\bibinfo {author} {\bibfnamefont {K.}~\bibnamefont
  {Laubscher}}, \bibinfo {author} {\bibfnamefont {D.}~\bibnamefont {Loss}}, \
  and\ \bibinfo {author} {\bibfnamefont {J.}~\bibnamefont {Klinovaja}},\
  }\bibfield  {title} {\enquote {\bibinfo {title} {{Majorana and parafermion
  corner states from two coupled sheets of bilayer graphene}}}, }\href
  {\doibase 10.1103/PhysRevResearch.2.013330} {\bibfield  {journal} {\bibinfo
  {journal} {Phys. Rev. Research}\ }\textbf {\bibinfo {volume} {2}},\ \bibinfo
  {pages} {013330} (\bibinfo {year} {2020})}\BibitemShut {NoStop}%
\bibitem [{\citenamefont {Hsu}\ \emph {et~al.}(2020)\citenamefont {Hsu},
  \citenamefont {Cole}, \citenamefont {Zhang},\ and\ \citenamefont
  {Sau}}]{HsuYT20PRL}%
  \BibitemOpen
  \bibfield  {author} {\bibinfo {author} {\bibfnamefont {Y.-T.}\ \bibnamefont
  {Hsu}}, \bibinfo {author} {\bibfnamefont {W.~S.}\ \bibnamefont {Cole}},
  \bibinfo {author} {\bibfnamefont {R.-X.}\ \bibnamefont {Zhang}}, \ and\
  \bibinfo {author} {\bibfnamefont {J.~D.}\ \bibnamefont {Sau}},\ }\bibfield
  {title} {\enquote {\bibinfo {title} {{Inversion-Protected Higher-Order
  Topological Superconductivity in Monolayer ${\mathrm{WTe}}_{2}$}}}, }\href
  {\doibase 10.1103/PhysRevLett.125.097001} {\bibfield  {journal} {\bibinfo
  {journal} {Phys. Rev. Lett.}\ }\textbf {\bibinfo {volume} {125}},\ \bibinfo
  {pages} {097001} (\bibinfo {year} {2020})}\BibitemShut {NoStop}%
\bibitem [{\citenamefont {{Wu}}\ \emph {et~al.}()\citenamefont {{Wu}},
  \citenamefont {{Liu}}, \citenamefont {{Thomale}},\ and\ \citenamefont
  {{Liu}}}]{XXWu19arXiv190510648W}%
  \BibitemOpen
  \bibfield  {author} {\bibinfo {author} {\bibfnamefont {X.}~\bibnamefont
  {{Wu}}}, \bibinfo {author} {\bibfnamefont {X.}~\bibnamefont {{Liu}}},
  \bibinfo {author} {\bibfnamefont {R.}~\bibnamefont {{Thomale}}}, \ and\
  \bibinfo {author} {\bibfnamefont {C.-X.}\ \bibnamefont {{Liu}}},\ }\bibfield
  {title} {\enquote {\bibinfo {title} {{High-$T_c$ Superconductor Fe(Se,Te)
  Monolayer: an Intrinsic, Scalable and Electrically-tunable Majorana
  Platform}}}, }\href@noop {} {\ }\Eprint {http://arxiv.org/abs/1905.10648}
  {arXiv:1905.10648} \BibitemShut {NoStop}%
\bibitem [{\citenamefont {Peng}(2020)}]{Peng20PRR}%
  \BibitemOpen
  \bibfield  {author} {\bibinfo {author} {\bibfnamefont {Y.}~\bibnamefont
  {Peng}},\ }\bibfield  {title} {\enquote {\bibinfo {title} {{Floquet
  higher-order topological insulators and superconductors with space-time
  symmetries}}}, }\href {\doibase 10.1103/PhysRevResearch.2.013124} {\bibfield
  {journal} {\bibinfo  {journal} {Phys. Rev. Research}\ }\textbf {\bibinfo
  {volume} {2}},\ \bibinfo {pages} {013124} (\bibinfo {year}
  {2020})}\BibitemShut {NoStop}%
\bibitem [{\citenamefont {Bomantara}\ and\ \citenamefont
  {Gong}(2020)}]{Bomantara20PRB}%
  \BibitemOpen
  \bibfield  {author} {\bibinfo {author} {\bibfnamefont {R.~W.}\ \bibnamefont
  {Bomantara}}\ and\ \bibinfo {author} {\bibfnamefont {J.}~\bibnamefont
  {Gong}},\ }\bibfield  {title} {\enquote {\bibinfo {title} {{Measurement-only
  quantum computation with Floquet Majorana corner modes}}}, }\href {\doibase
  10.1103/PhysRevB.101.085401} {\bibfield  {journal} {\bibinfo  {journal}
  {Phys. Rev. B}\ }\textbf {\bibinfo {volume} {101}},\ \bibinfo {pages}
  {085401} (\bibinfo {year} {2020})}\BibitemShut {NoStop}%
\bibitem [{Sup()}]{SuppInf}%
  \BibitemOpen
  \href@noop {} {\bibinfo  {journal} {{See the Supplemental Material for
  details}}\ }\BibitemShut {NoStop}%
\bibitem [{\citenamefont {Schnyder}\ \emph {et~al.}(2008)\citenamefont
  {Schnyder}, \citenamefont {Ryu}, \citenamefont {Furusaki},\ and\
  \citenamefont {Ludwig}}]{Schnyder_PRB_2008}%
  \BibitemOpen
\bibfield  {journal} {  }\bibfield  {author} {\bibinfo {author} {\bibfnamefont
  {A.~P.}\ \bibnamefont {Schnyder}}, \bibinfo {author} {\bibfnamefont
  {S.}~\bibnamefont {Ryu}}, \bibinfo {author} {\bibfnamefont {A.}~\bibnamefont
  {Furusaki}}, \ and\ \bibinfo {author} {\bibfnamefont {A.~W.~W.}\ \bibnamefont
  {Ludwig}},\ }\bibfield  {title} {\enquote {\bibinfo {title} {Classification
  of topological insulators and superconductors in three spatial dimensions}},
  }\href {\doibase 10.1103/PhysRevB.78.195125} {\bibfield  {journal} {\bibinfo
  {journal} {Phys. Rev. B}\ }\textbf {\bibinfo {volume} {78}},\ \bibinfo
  {pages} {195125} (\bibinfo {year} {2008})}\BibitemShut {NoStop}%
\bibitem [{foo()}]{footnote}%
  \BibitemOpen
  \href@noop {} {}\bibinfo {note} {We refer to the boundary states that have
  the same origin like the hinge boundary states as "hinge Majorana arcs" on a
  2D surface.}\BibitemShut {Stop}%
\bibitem [{\citenamefont {Kitaev}(2001)}]{Kitaev_2001}%
  \BibitemOpen
  \bibfield  {author} {\bibinfo {author} {\bibfnamefont {A.~Y.}\ \bibnamefont
  {Kitaev}},\ }\bibfield  {title} {\enquote {\bibinfo {title} {{Unpaired
  Majorana fermions in quantum wires}}}, }\href {\doibase
  10.1070/1063-7869/44/10s/s29} {\bibfield  {journal} {\bibinfo  {journal}
  {Physics-Uspekhi}\ }\textbf {\bibinfo {volume} {44}},\ \bibinfo {pages} {131}
  (\bibinfo {year} {2001})}\BibitemShut {NoStop}%
\bibitem [{\citenamefont {Alicea}(2012)}]{Alicea_2012}%
  \BibitemOpen
  \bibfield  {author} {\bibinfo {author} {\bibfnamefont {J.}~\bibnamefont
  {Alicea}},\ }\bibfield  {title} {\enquote {\bibinfo {title} {New directions
  in the pursuit of majorana fermions in solid state systems}}, }\href
  {\doibase 10.1088/0034-4885/75/7/076501} {\bibfield  {journal} {\bibinfo
  {journal} {Reports on Progress in Physics}\ }\textbf {\bibinfo {volume}
  {75}},\ \bibinfo {pages} {076501} (\bibinfo {year} {2012})}\BibitemShut
  {NoStop}%
\bibitem [{\citenamefont {Mourik}\ \emph {et~al.}(2012)\citenamefont {Mourik},
  \citenamefont {Zuo}, \citenamefont {Frolov}, \citenamefont {Plissard},
  \citenamefont {Bakkers},\ and\ \citenamefont {Kouwenhoven}}]{Mourik1003}%
  \BibitemOpen
  \bibfield  {author} {\bibinfo {author} {\bibfnamefont {V.}~\bibnamefont
  {Mourik}}, \bibinfo {author} {\bibfnamefont {K.}~\bibnamefont {Zuo}},
  \bibinfo {author} {\bibfnamefont {S.~M.}\ \bibnamefont {Frolov}}, \bibinfo
  {author} {\bibfnamefont {S.~R.}\ \bibnamefont {Plissard}}, \bibinfo {author}
  {\bibfnamefont {E.~P. A.~M.}\ \bibnamefont {Bakkers}}, \ and\ \bibinfo
  {author} {\bibfnamefont {L.~P.}\ \bibnamefont {Kouwenhoven}},\ }\bibfield
  {title} {\enquote {\bibinfo {title} {{Signatures of Majorana Fermions in
  Hybrid Superconductor-Semiconductor Nanowire Devices}}}, }\href {\doibase
  10.1126/science.1222360} {\bibfield  {journal} {\bibinfo  {journal}
  {Science}\ }\textbf {\bibinfo {volume} {336}},\ \bibinfo {pages} {1003}
  (\bibinfo {year} {2012})}\BibitemShut {NoStop}%
\bibitem [{\citenamefont {Lutchyn}\ \emph {et~al.}(2010)\citenamefont
  {Lutchyn}, \citenamefont {Sau},\ and\ \citenamefont
  {Das~Sarma}}]{Lutchyn_PRL_2010}%
  \BibitemOpen
  \bibfield  {author} {\bibinfo {author} {\bibfnamefont {R.~M.}\ \bibnamefont
  {Lutchyn}}, \bibinfo {author} {\bibfnamefont {J.~D.}\ \bibnamefont {Sau}}, \
  and\ \bibinfo {author} {\bibfnamefont {S.}~\bibnamefont {Das~Sarma}},\
  }\bibfield  {title} {\enquote {\bibinfo {title} {Majorana fermions and a
  topological phase transition in semiconductor-superconductor
  heterostructures}}, }\href {\doibase 10.1103/PhysRevLett.105.077001}
  {\bibfield  {journal} {\bibinfo  {journal} {Phys. Rev. Lett.}\ }\textbf
  {\bibinfo {volume} {105}},\ \bibinfo {pages} {077001} (\bibinfo {year}
  {2010})}\BibitemShut {NoStop}%
\bibitem [{\citenamefont {Aasen}\ \emph {et~al.}(2016)\citenamefont {Aasen},
  \citenamefont {Hell}, \citenamefont {Mishmash}, \citenamefont {Higginbotham},
  \citenamefont {Danon}, \citenamefont {Leijnse}, \citenamefont {Jespersen},
  \citenamefont {Folk}, \citenamefont {Marcus}, \citenamefont {Flensberg},\
  and\ \citenamefont {Alicea}}]{Aasen_2016_PRX}%
  \BibitemOpen
  \bibfield  {author} {\bibinfo {author} {\bibfnamefont {D.}~\bibnamefont
  {Aasen}}, \bibinfo {author} {\bibfnamefont {M.}~\bibnamefont {Hell}},
  \bibinfo {author} {\bibfnamefont {R.~V.}\ \bibnamefont {Mishmash}}, \bibinfo
  {author} {\bibfnamefont {A.}~\bibnamefont {Higginbotham}}, \bibinfo {author}
  {\bibfnamefont {J.}~\bibnamefont {Danon}}, \bibinfo {author} {\bibfnamefont
  {M.}~\bibnamefont {Leijnse}},  \emph {et~al.},\ }\bibfield  {title} {\enquote
  {\bibinfo {title} {{Milestones Toward Majorana-Based Quantum Computing}}},
  }\href {\doibase 10.1103/PhysRevX.6.031016} {\bibfield  {journal} {\bibinfo
  {journal} {Phys. Rev. X}\ }\textbf {\bibinfo {volume} {6}},\ \bibinfo {pages}
  {031016} (\bibinfo {year} {2016})}\BibitemShut {NoStop}%
\bibitem [{\citenamefont {Sun}\ \emph {et~al.}(2016)\citenamefont {Sun},
  \citenamefont {Zhang}, \citenamefont {Hu}, \citenamefont {Li}, \citenamefont
  {Wang}, \citenamefont {Ma}, \citenamefont {Xu}, \citenamefont {Gao},
  \citenamefont {Guan}, \citenamefont {Li}, \citenamefont {Liu}, \citenamefont
  {Qian}, \citenamefont {Zhou}, \citenamefont {Fu}, \citenamefont {Li},
  \citenamefont {Zhang},\ and\ \citenamefont {Jia}}]{sun_PRL_2016}%
  \BibitemOpen
  \bibfield  {author} {\bibinfo {author} {\bibfnamefont {H.-H.}\ \bibnamefont
  {Sun}}, \bibinfo {author} {\bibfnamefont {K.-W.}\ \bibnamefont {Zhang}},
  \bibinfo {author} {\bibfnamefont {L.-H.}\ \bibnamefont {Hu}}, \bibinfo
  {author} {\bibfnamefont {C.}~\bibnamefont {Li}}, \bibinfo {author}
  {\bibfnamefont {G.-Y.}\ \bibnamefont {Wang}}, \bibinfo {author}
  {\bibfnamefont {H.-Y.}\ \bibnamefont {Ma}},  \emph {et~al.},\ }\bibfield
  {title} {\enquote {\bibinfo {title} {{Majorana Zero Mode Detected with Spin
  Selective Andreev Reflection in the Vortex of a Topological
  Superconductor}}}, }\href {\doibase 10.1103/PhysRevLett.116.257003}
  {\bibfield  {journal} {\bibinfo  {journal} {Phys. Rev. Lett.}\ }\textbf
  {\bibinfo {volume} {116}},\ \bibinfo {pages} {257003} (\bibinfo {year}
  {2016})}\BibitemShut {NoStop}%
\bibitem [{\citenamefont {Sticlet}\ \emph {et~al.}(2012)\citenamefont
  {Sticlet}, \citenamefont {Bena},\ and\ \citenamefont {Simon}}]{Sticlet12PRL}%
  \BibitemOpen
  \bibfield  {author} {\bibinfo {author} {\bibfnamefont {D.}~\bibnamefont
  {Sticlet}}, \bibinfo {author} {\bibfnamefont {C.}~\bibnamefont {Bena}}, \
  and\ \bibinfo {author} {\bibfnamefont {P.}~\bibnamefont {Simon}},\ }\bibfield
   {title} {\enquote {\bibinfo {title} {{Spin and Majorana Polarization in
  Topological Superconducting Wires}}}, }\href {\doibase
  10.1103/PhysRevLett.108.096802} {\bibfield  {journal} {\bibinfo  {journal}
  {Phys. Rev. Lett.}\ }\textbf {\bibinfo {volume} {108}},\ \bibinfo {pages}
  {096802} (\bibinfo {year} {2012})}\BibitemShut {NoStop}%
\bibitem [{\citenamefont {K\"onig}\ \emph {et~al.}(2008)\citenamefont
  {K\"onig}, \citenamefont {Buhmann}, \citenamefont {W.~Molenkamp},
  \citenamefont {Hughes}, \citenamefont {Liu}, \citenamefont {Qi},\ and\
  \citenamefont {Zhang}}]{Konig08JPSJ}%
  \BibitemOpen
  \bibfield  {author} {\bibinfo {author} {\bibfnamefont {M.}~\bibnamefont
  {K\"onig}}, \bibinfo {author} {\bibfnamefont {H.}~\bibnamefont {Buhmann}},
  \bibinfo {author} {\bibfnamefont {L.}~\bibnamefont {W.~Molenkamp}}, \bibinfo
  {author} {\bibfnamefont {T.}~\bibnamefont {Hughes}}, \bibinfo {author}
  {\bibfnamefont {C.-X.}\ \bibnamefont {Liu}}, \bibinfo {author} {\bibfnamefont
  {X.-L.}\ \bibnamefont {Qi}}, \ and\ \bibinfo {author} {\bibfnamefont {S.-C.}\
  \bibnamefont {Zhang}},\ }\bibfield  {title} {\enquote {\bibinfo {title} {{The
  Quantum Spin Hall Effect: Theory and Experiment}}}, }\href {\doibase
  10.1143/JPSJ.77.031007} {\bibfield  {journal} {\bibinfo  {journal} {Journal
  of the Physical Society of Japan}\ }\textbf {\bibinfo {volume} {77}},\
  \bibinfo {pages} {031007} (\bibinfo {year} {2008})}\BibitemShut {NoStop}%
\bibitem [{\citenamefont {Tan}\ \emph {et~al.}(2019)\citenamefont {Tan},
  \citenamefont {Zhao}, \citenamefont {Liu}, \citenamefont {Xue}, \citenamefont
  {Yu}, \citenamefont {Wang},\ and\ \citenamefont
  {Yu}}]{Tan_PRL_2019_Simulation}%
  \BibitemOpen
  \bibfield  {author} {\bibinfo {author} {\bibfnamefont {X.}~\bibnamefont
  {Tan}}, \bibinfo {author} {\bibfnamefont {Y.~X.}\ \bibnamefont {Zhao}},
  \bibinfo {author} {\bibfnamefont {Q.}~\bibnamefont {Liu}}, \bibinfo {author}
  {\bibfnamefont {G.}~\bibnamefont {Xue}}, \bibinfo {author} {\bibfnamefont
  {H.-F.}\ \bibnamefont {Yu}}, \bibinfo {author} {\bibfnamefont {Z.~D.}\
  \bibnamefont {Wang}}, \ and\ \bibinfo {author} {\bibfnamefont
  {Y.}~\bibnamefont {Yu}},\ }\bibfield  {title} {\enquote {\bibinfo {title}
  {{Simulation and Manipulation of Tunable Weyl-Semimetal Bands Using
  Superconducting Quantum Circuits}}}, }\href {\doibase
  10.1103/PhysRevLett.122.010501} {\bibfield  {journal} {\bibinfo  {journal}
  {Phys. Rev. Lett.}\ }\textbf {\bibinfo {volume} {122}},\ \bibinfo {pages}
  {010501} (\bibinfo {year} {2019})}\BibitemShut {NoStop}%
\end{thebibliography}%

\end{document}